\RequirePackage{lineno}
\documentclass[aps,prl,twocolumn,floatfix,reprint]{revtex4-1}
\usepackage{graphicx}
\usepackage{color}
\usepackage{amssymb} 
\usepackage{amsmath} 
\usepackage{natbib}
\usepackage{soul}
\usepackage{titlesec}
\usepackage{physics}
\usepackage[utf8]{inputenc}
\usepackage{hyperref}
\hypersetup{
    colorlinks=true,
    linkcolor=blue,
    filecolor=magenta,      
    urlcolor=cyan,
    citecolor=blue,
}

\setcitestyle{super, numbers, square,}

\usepackage{color}
\definecolor{pu}{rgb}{0,0.,0.55}  
\definecolor{rd}{rgb}{1,0,0} 
\definecolor{gr}{rgb}{0.6,0.6,0.6}

\begin{document}
\title{Valley interference and spin exchange at the atomic scale in silicon}

\author{B. Voisin$^{1*}$}
\author{J. Bocquel$^{1*}$}
\author{A. Tankasala$^{2*}$}
\author{M. Usman$^{3,4*}$}
\author{J. Salfi$^1$}
\author{R. Rahman$^{2,5}$}
\author{M.Y. Simmons$^1$}
\author{L.C.L. Hollenberg$^3$}
\author{S. Rogge$^1$}

\affiliation{{$^1$\, Centre for Quantum Computation and Communication Technology, School of Physics, The University of New South Wales, Sydney, 2052, NSW, Australia} \\
{$^2$\, Electrical and Computer Engineering Department, Purdue University, West Lafayette, Indiana, USA}\\
{$^3$\, Centre for Quantum Computation and Communication Technology, School of Physics, The University of Melbourne, Parkville, 3010, VIC, Australia}\\
{$^4$\, School of Computing and Information Systems, Melbourne, School of Engineering, The University of Melbourne, Parkville, Victoria 3010, Australia.}\\
{$^5$\, School of Physics, The University of New South Wales, Sydney, 2052, NSW, Australia} \\
{$^{*}$\,These authors contributed equally}}

\begin{abstract}
Tunneling is a fundamental quantum process with no classical equivalent, which can compete with Coulomb interactions to give rise to complex phenomena. Phosphorus dopants in silicon can be placed with atomic precision to address the different regimes arising from this competition. However, they exploit wavefunctions relying on crystal band symmetries, which tunneling interactions are inherently sensitive to. Here we directly image lattice-aperiodic valley interference between coupled atoms in silicon using scanning tunneling microscopy. Our atomistic analysis unveils the role of envelope anisotropy, valley interference and dopant placement on the Heisenberg spin exchange interaction. We find that the exchange can become immune to valley interference by engineering in-plane dopant placement along specific crystallographic directions. A vacuum-like behaviour is recovered, where the exchange is maximised to the overlap between the donor orbitals, and pair-to-pair variations limited to a factor of less than 10 considering the accuracy in dopant positioning. This robustness remains over a large range of distances, from the strongly Coulomb interacting regime relevant for high-fidelity quantum computation to strongly coupled donor arrays of interest for quantum simulation in silicon.
\end{abstract}

\maketitle
\section{Introduction}
Quantum tunneling is a widespread phenomena, where a wavefunction leaking as an evanescent mode can be transmitted through a finite barrier. This simple description applies to many natural systems, accounting for instance for molecular conformation or radioactivity~\cite{Razavy2003}. In solid-state systems, nanoscale structures can use tunnelling effects in novel functionalities developed for CMOS electronics~\cite{Ionescu2011} as well as for quantum devices, where electrons can be confined and manipulated in quantum information schemes~\cite{Hanson2007,Hensgens2017,Fuechsle2012,He2019}.

For many-body quantum devices, the tunnel interaction $t$, which can couple the different sites, usually competes with the on-site interaction $U$ (also called the charging energy). This competition is core to the so-called Fermi-Hubbard model. For fermions, this model needs to be discussed together with the Pauli exclusion principle, which forbids to form a state with two parallel spins on the same orbital~\cite{Salfi2016,Mazurenko2017}. When the Coulomb interactions $U$ dominate over $t$, the system maps to the Heisenberg model with an exchange interaction which directly links to the tunnel coupling through the relationship $J=4t^2/U$. This Heisenberg limit is favourable for quantum simulations of magnetism~\cite{Hensgens2017, Dehollain2020} and for fast quantum computation schemes~\cite{Kane1998,Russ2016,He2019}. The mapping to the Heisenberg model breaks down when $t$ approaches $U$ and when the system is brought away from half-filling, to give rise to a rich phase diagram with links to exotic superconductivity and spin liquids~\cite{Georgescu2014,Balents2010,Gull2013,Salfi2016,Mazurenko2017}.

\begin{figure*}[htp]
\begin{center}
\includegraphics[width=18cm]{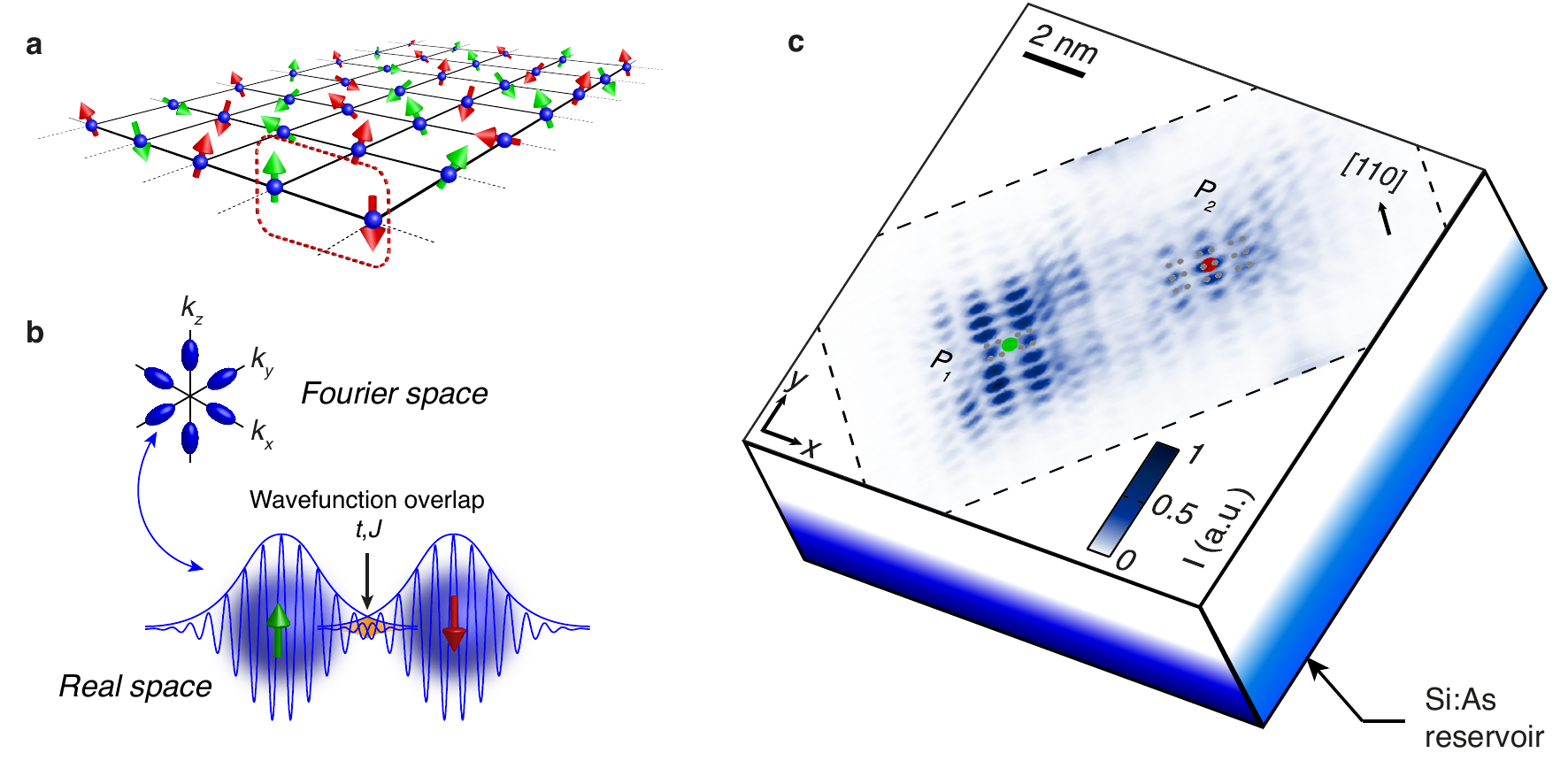}
\caption{\textbf{Direct measurement of coupled donors in silicon.} \textbf{a} Schematics of a 2D-array of single electron spins bound to phosphorus atoms in $\rm{^{28}Si}$, where two-qubit operations can occur between nearest neighbours, i.e. red/green spin pairs, using the Heisenberg exchange interaction. \textbf{b} (top) A donor's electron wavefunction oscillates at the valley wavevector $\bf{k_{\mu}}\sim\rm{0.81}\bf{k_0}$, with $k_0{=}2\pi/a_0$. Silicon presents a mass anisotropy which results in the donor orbital envelope to be anisotropic as well, as highlighted by their ovoidal shape in Fourier space. (bottom) Tunnel and exchange interactions result from the overlap between the two donor orbitals, represented by the yellow overlap area, which is sensitive to both envelope decay and valley interference between donors. \textbf{c} Experimental real-space map of the quasi-particle wave function of an exchange-coupled donor pair's two-electron neutral state. Sequential transport occurs vertically from a highly doped Si:As substrate, which acts as an electron reservoir, to a donor pair found in a low-doped phosphorus $\delta$-layer, and then to the STM tip above (not shown). The red and green dots represent the surface projections of the pinpointed lattice positions of the two donors. The two donors are separated by ${6.5a_0\sqrt{2}}$ along [110] (perpendicular to the Si dimers, ${a_0}$ is the silicon lattice constant), ${0.5a_0\sqrt{2}/2}$ (parallel to the Si dimers) along ${[1\bar{1}0]}$, and by ${1.25\,a_0}$ in depth. The grey dots represent the silicon atom positions of the $2 \times 1$ reconstructed surface. The black dashed contour indicates the experimentally measured area.}
\label{fig1}
\end{center}
\end{figure*}

Phosphorus donor-bound spins in silicon are highly suitable to explore and harness these different regimes. Donors can be placed at nanometer scale distances from each other in the silicon crystal~\cite{Dupont-Ferrier2013,Gonzalez-Zalba2014,Dehollain2014,Weber2014,Broome2018}, where direct tunneling interactions dominate over dipolar coupling. Notably, 2D donor arrays (see Fig.~\ref{fig1}a) can be fabricated using scanning tunneling lithography~\cite{Fuechsle2012,Weber2014}, where the atoms can be placed anywhere in a single atomic plane. These atomically precise devices can be engineered to achieve both the Heisenberg limit~\cite{He2019} or the non-perturbative tunneling interactions regime at short inter-dopant distances~\cite{Salfi2016}, with ratios $U/t$ possibly lower than 10.

The direct relationship between tunnel and exchange coupling can be understood conceptually as they are both based on wavefunction overlap. Hence, they are sensitive to the same physical effects, and these atomic-scale wavefunctions must be precisely investigated to warrant the development of applications which requires the exchange interaction to be well controlled. In contrast to atoms in the vacuum~\cite{Mazurenko2017}, donor-bound electrons in silicon acquire properties of the crystal band structure. In particular, silicon is an indirect band gap semiconductor~\cite{Gunawan2006,Randeria2018,Wang2018}, with the presence of an anisotropic mass and a valley degree of freedom~\cite{Kohn1955,Salfi2014}. More precisely, the finite valley momentum is aperiodic with the silicon lattice, a feature which can also be found in 2D material structures~\cite{Mudd2016,Gonzalez2016,Movva2018}. This aperiodicity causes interference between oscillating valley states from different donors (see Fig.~\ref{fig1}b). Consequently, the tunnel and exchange interactions can vary strongly with small lattice site variations in the donor positions, as they induce changes in both valley interference condition and envelope overlap. Such reduction of the exchange energy compared to hydrogenic systems has been observed in ensemble work~\cite{Cullis1970,Andres1981}, but a direct link to valley interference cannot be accessed from any ensemble measurement. Understanding the impact of valley interference at the atomic scale, i.e. at the wavefunction level, has become essential in the context of quantum computing~\cite{Kane1998,Koiller2001}, but the predicted valley-induced exchange variations range within 5 orders of magnitude~\cite{Kane1998,Koiller2001,Wellard2005,Testolin2007,Rahman2011,Saraiva2015,Gamble2015,Wang2016,Song2016}, which makes it challenging to design scalable donor qubit architectures with uniform couplings. The sensitivity of both tunnel and exchange interactions to the precise donor positions makes it essential to verify experimentally the presence and magnitude of valley interference. This is challenging to achieve in transport experiments~\cite{Dupont-Ferrier2013,Gonzalez-Zalba2014,Dehollain2014,Weber2014,Broome2018,He2019}, which are unable to discriminate between valley interference and envelope effects. Here, we instead implement a real-space probe into coupled donor wavefunctions. This experimental approach leads us to detail the precise role of envelope anisotropy, crystal symmetries, and dopant placement accuracy to reconcile the apparent disparity in the exchange variations predictions. In particular, we consider the dopant placement precision which can be ultimately achieved by scanning probe lithography~\cite{Fuechsle2012}. We find that exchange variations from pair to pair can be minimised to within an order of magnitude and investigate the potential of this strategy in view of quantum processing using exchange-based donor arrays.
 
\begin{figure*}[htp]
\begin{center}
\includegraphics[width=18cm]{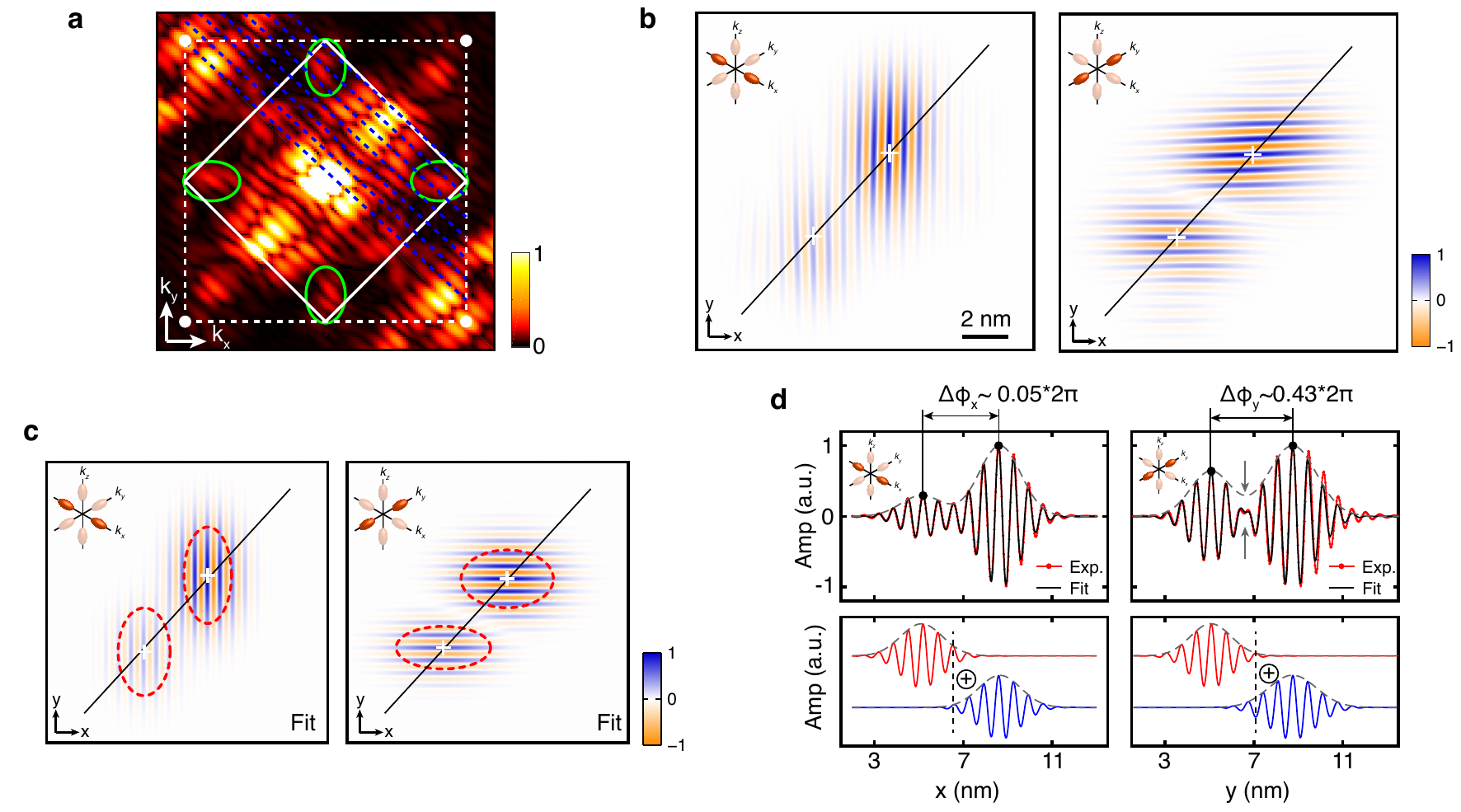}
\caption{\textbf{Visualizing and quantifying valley interference between coupled donors. Pair \#1. a} 2D FFT of the STM image, centred on the first Brillouin zone (solid white lines). The white dots are located at $\{\pm k_0,\pm k_0\}$. The FFT shows a clear valley signal around ${k_{\mu}\sim0.81k_0}$ evidenced by the green ellipsoids.  The FFT also shows diagonal slices (blue dotted lines), cutting through the valley components of the FFT, which evidence the geometric interference between the two donors. \textbf{b} Real-space images of the valley interference, obtained after inverse Fourier transform of the FFT filtered around ${k_x\sim0.81k_0}$ for the $x$-valley interference (left) or ${k_y\sim0.81k_0}$ for the $y$-valley interference (right). The white crosses indicate the donor positions. The $x$-valleys look in-phase with the vertical stripes running continuously from one donor to the other. The $y$-valleys look out-of-phase with a clear discontinuity in the oscillatory valley pattern between the two donors. \textbf{c} 2D fits of the valley images, from which the position of the valley maxima and of the valley frequency, and hence the valley phase differences $\Delta\phi_x$ and $\Delta\phi_y$, can be obtained. The red dashed ellipsoids correspond to the donor envelope part of the fits, which highlights their anisotropy. \textbf{d} Line cuts taken through the ion-ion direction of both valley images (red lines) and their fit (black lines), for the $x$-valleys (left) and the $y$-valleys (right). The grey dashed lines represent the envelope part of the fits only. The $x$-valleys are in-phase, which results in the maxima of valley signal to always reach the envelop part. On the contrary, the $y$-valley are out-of-phase and the grey arrows in-between the two donors highlight the clear reduction of the valley signal compared to the envelope part as a result of the destructive interference.}
\label{fig2}
\end{center}
\end{figure*}

\section{Results}
\subsection{Valley interference and anisotropic envelope in exchange-coupled donors}

In order to directly probe valley interference between donors, we have designed and fabricated samples to host isolated donor pairs embedded beneath a silicon surface (see Methods). Spatially resolved transport is performed at 4\,K between a heavily doped reservoir and the tip of a scanning tunneling microscope (STM). An STM image of a donor pair is shown in Fig~\ref{fig1}c. It is taken at a tip-sample bias of $V_b=-0.95$\,V, close to the zero electric-field condition where the tip does not influence the two-electron neutral molecular state~\cite{Salfi2016,Salfi2018}. More specifically, the spatially resolved tunneling current to the tip represents a quasi-particle wavefunction (QPWF), corresponding to the sum of the transitions from the two-electron ground state to energetically available one-electron states. Hence, the STM image contains two-electron wavefunction information including interference between the two donor wavefunctions~\cite{Rontani2005,Salfi2016,Salfi2018}. Moreover, the exact site positions of the two donor ions ($P_1$ and $P_2$, respectively green and red dot) are determined out of different possible position configurations using a comprehensive image symmetry recognition protocol~\cite{Usman2016,Brazdova2017}, which notably include the tip orbital, as the oscillating pattern observed for each donor qualitatively varies depending on their position in the silicon lattice. In particular, donor $P_1$ is located in the $z{=}5.5\,a_0$ atomic plane, in-between two dimer rows, and presents a minima at the surface projection of the ion location. The image corresponding to donor $P_1$ presents two rows of $2{\times}3$ local maxima running along the $[1\bar{1}0]$ crystallographic axis, on each side of the ion location. On the contrary, donor $P_2$ is found to be in the $z{=}6.75\,a_0$ atomic plane, and sits directly underneath the middle of a dimer row of the $2{\times}1$ reconstructed surface. Donor $P_2$ presents a maximum (instead of a minimum for $P_1$) at this location, part of a row of three local maxima running along this dimer.

\begin{figure}[htp]
\includegraphics[width=8.8cm]{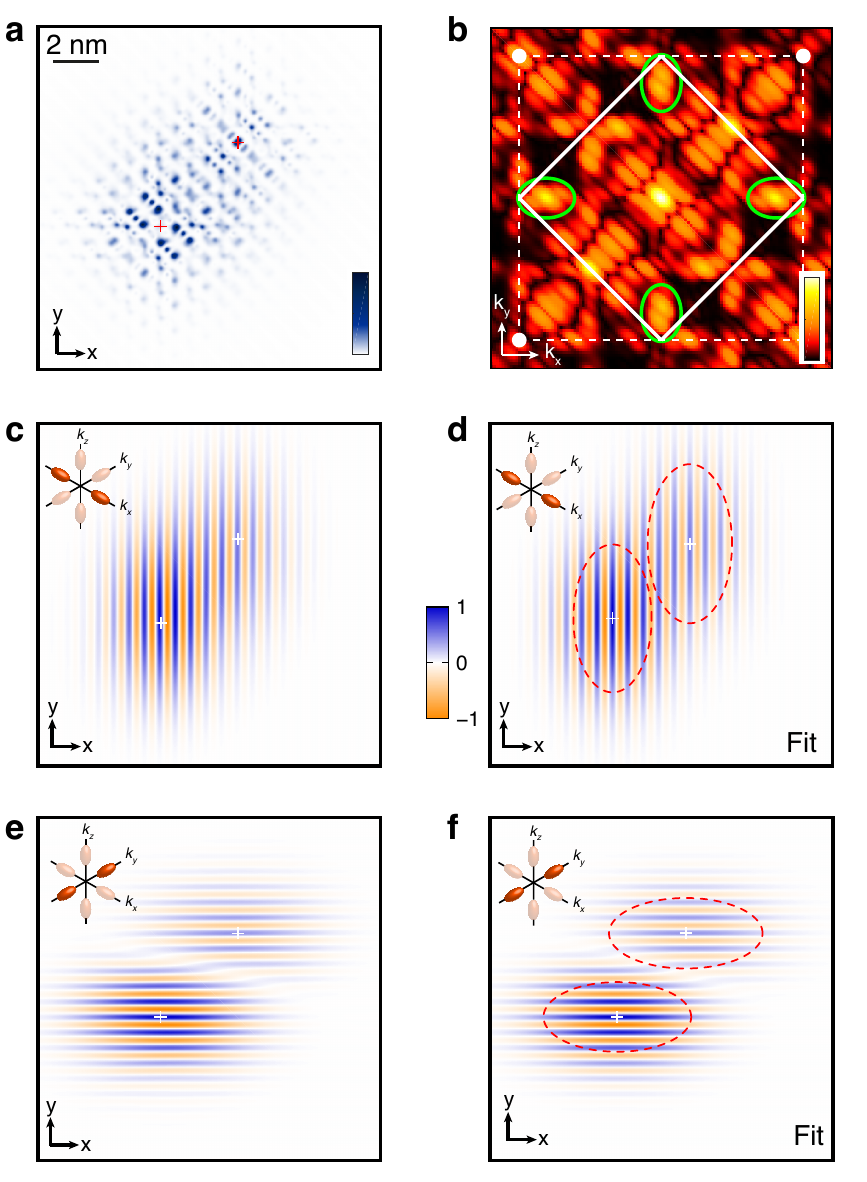}
\caption{\textbf{Theoretical two-electron STM image and valley interference.} \textbf{a} Theoretical STM image computed using the pinpointed locations of the two donors and FCI calculations performed to determine the two-electron wavefunctions based on tight-binding one-electron states. \textbf{b} 2D Fourier transform of the STM image focussed on the first Brillouin zone. The valley components at 0.81$k_0$ and the diagonal slices highlighting the interference are clearly visible. \textbf{c} Real-space image of the $x$-valley obtained by filtering the FFT around ${k_x\sim0.81k_0}$. \textbf{d} Corresponding fitted image, using the same fitting procedure as for the experimental data, and from which $\Delta\phi_x$ and the ratio $b/a$ for each donor (red dashed ellipsoids) can be extracted. As for the experiment, the $x$-valleys are found to be in-phase. \textbf{e,f} Same for the $y$-valleys, which like for the experiment are out-of-phase.}
\label{fig2bis}
\end{figure}

\begin{figure}[htp!]
\begin{center}
\includegraphics[width=8.8cm]{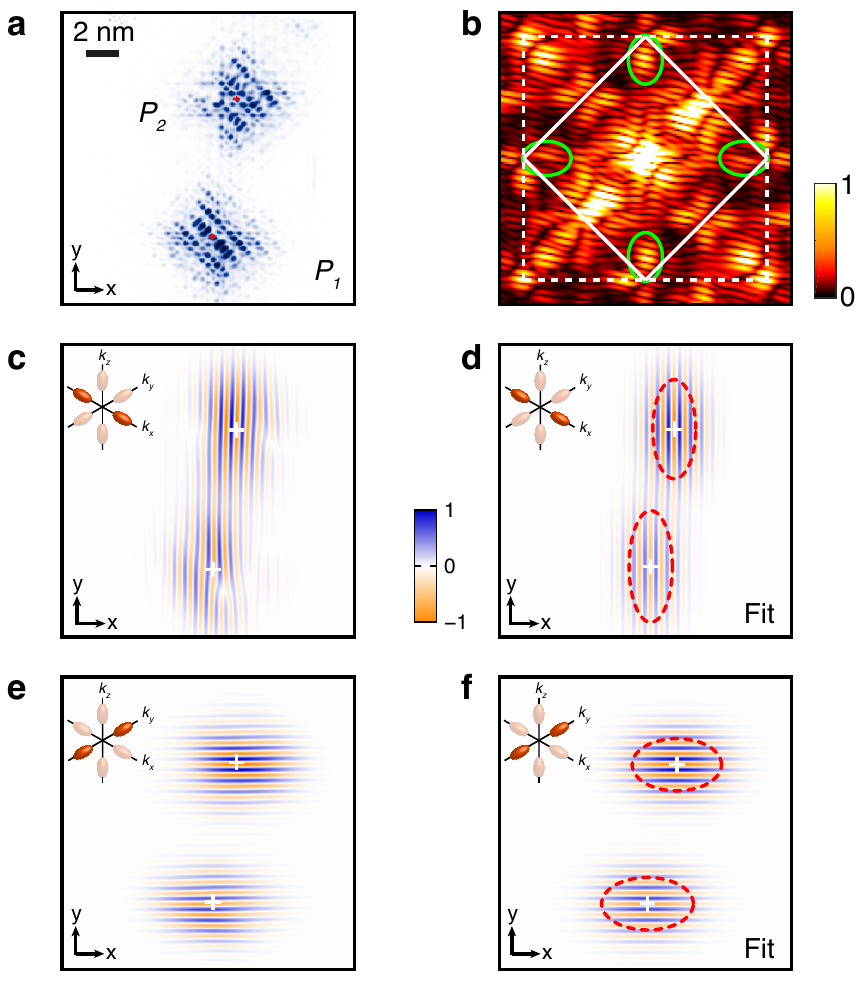}
\caption{\textbf{Valley interference for pair \#2.} \textbf{a} Experimental STM image of pair \#2. The red crosses indicate the ion locations. \textbf{b} FFT of the STM image, centred on the first Brillouin zone. As for pair\#1, the FFT shows a valley signal at 0.81$k_0$ (green ellipsoids) and diagonal slices that indicate the interference, whose separation and slope is linked  to the inter-donor distance and orientation. \textbf{c} Real-space image of the $x$-valley obtained by filtering the FFT around ${k_x\sim0.81k_0}$, same procedure as for pair \#1. Same scale as the STM image shown in \textbf{a}. The observed distortions in the $x$-valley image, notably on the right-hand side of $P_1$, are attributed to an instability of the tunneling tip during the measurement. \textbf{d} Corresponding fitted image, using the same fitting procedure as for pair \#1, and from which $\Delta\phi_x$ and the ratio $b/a$ for each donor (red dotted ellipsoids) can be extracted. \textbf{e,f} Same for the $y$-valleys.}
\label{fig2bis2}
\end{center}
\end{figure}

\begin{figure}[htp!]
\begin{center}
\includegraphics[width=8.8cm]{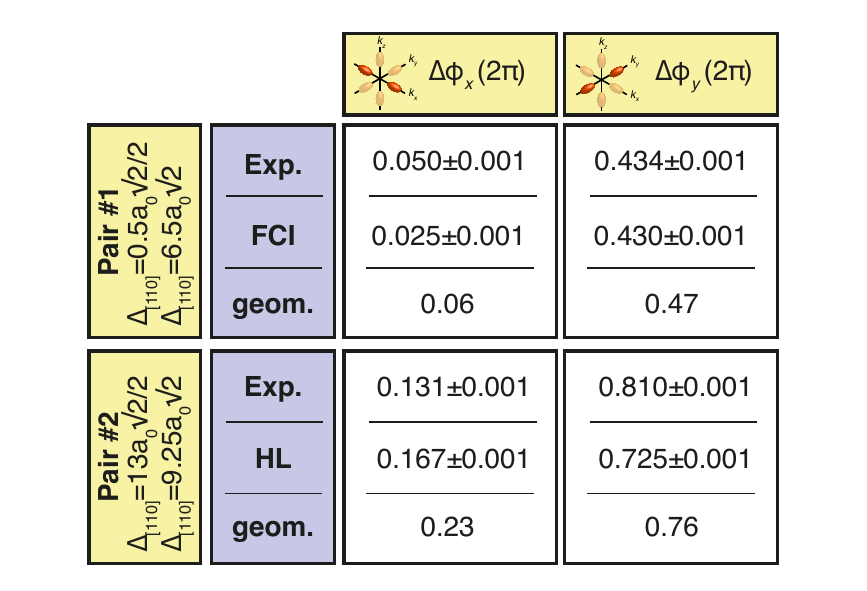}
\caption{\textbf{Table summarising the valley phase differences obtained for the two pairs.} The $x$ and $y$-valley phase difference for each pair obtained from fitting the 2D valley images can be compared to the geometrical phase differences obtained from the pinpointed lattice separation between the two donors along [110] and $[1\bar{1}0]$ and assuming a valley momentum at $0.81k_0$.}
\label{fig2sum}
\end{center}
\end{figure}

The reduced symmetries of the STM image compared to a spherical $1s$-orbital for atoms in the vacuum, and the alignment of its local maxima with the dimer rows at the silicon surface clearly indicate the presence of both lattice and valley frequencies in the donor wavefunctions. These frequencies can be probed in more detail in Fourier space~\cite{Saraiva2016}. A contour mask was first applied around the two donors to avoid the presence of spurious frequencies in the fast Fourier transform (FFT) from the image boundaries. The first Brillouin zone of the resulting Fourier transform is shown in Fig.~\ref{fig2}a. The different Fourier components which can be observed correspond to scattering processes between valley and lattice wavevectors~\cite{Saraiva2016}. In particular, the frequencies of the silicon $2\times1$ surface reconstruction can be identified at $(k_x,k_y)=\{\pm1,\pm1\}k_0$ (with $k_0{=}2\pi/a_0$), corresponding to the $a_0\sqrt{2}$ periodicity between the dimer rows, i.e. along $[110]$, as well $(k_x,k_y)=\pm(0.5,0.5)k_0$ corresponding to the $a_0\sqrt{2}/2$ periodicity along [110] between each pair of silicon atoms forming a dimer. The components of interest for the valley interference analysis are found around $\pm(0.81,0)k_0$ and $\pm(0,0.81)k_0$. They relate to valley scattering processes, respectively between the $\pm x$ and $\pm z$-valleys, and the $\pm y$ and $\pm z$-valleys. It is important to note that these valley processes can be completely dissociated from any lattice frequency-associated process, since they rely on the valley momentum only. Moreover, the Fourier transform presents clear diagonal slices (blue dashed lines), which match the geometric destructive interference condition between the two donors. Their position with respect to the valley signal gives a hint to whether the valley states are in or out-of-phase, if their corresponding signal in the FFT falls in-between or on one these diagonal stripes, respectively (See Supplementary Note 1). In order to focus on these valley interference processes, a Gaussian filter was applied to the FFT around the valley signal (see green ellipsoids in Fig.~\ref{fig2}a) and transformed back to real space. The left image in Fig.~\ref{fig2}b shows the $x$-component result of this filtering: a set of vertical stripes, corresponding to the valley oscillations in the $x$-direction can be observed for each donor, with the valley phase pinned at each donor site. For this particular inter-donor distance, the $x$-valleys look in-phase as the vertical stripes run continuously from one donor to the other. The same filtering procedure was performed for the $y$-valley real-space image. In contrast to the $x$-valley, the $y$-valleys look out of phase with a clear break in the continuity of the horizontal stripes in-between the donors and a succession of phase slips. In order to be more quantitative, these valley images can be fitted to the sum of two 2D envelopes oscillating at the same frequency. The fitted 2D images are shown in Fig~\ref{fig2}c. They give the same visual impression of the fits as that of the experimental images. Furthermore, the valley phase differences can be extracted from the fits, yielding $k_x=(0.8077\pm0.0001)^*k_0$, $k_y=(0.8039\pm0.0001)^*k_0$, $\Delta\phi_x=(0.050\pm0.001)^*2\pi$ and $\Delta\phi_y=(0.434\pm0.001)^*2\pi$, which confirms that the $x$-valleys are in-phase and that the $y$-valleys are out-of-phase. We show in Fig.~\ref{fig2}d line cuts through the two ions of each valley image, experimental ones and their respective fit, to notably highlight the clear reduction of the $y$-valley signal in-between the two donors compared to the sum of the envelopes of the two donors due to the destructive interference.

The procedure developed here establishes the existence of a geometric valley phase interference between donors, which depends on their relative position. This phase interference is by construction included in the Heitler-London regime of vanishing tunnel coupling, as the two electrons fully localise on the donors~\cite{Salfi2016}. In this regime, the STM image becomes equivalent to the sum of the charge distributions of two independent single donors (see Fig.~\ref{fig2}d and Supplementary Note 1). However, in our case the inter-donor distance is around 5\,nm, at which some deviation from the Heitler-London regime is expected~\cite{Rahman2011,Saraiva2015}, as highlighted by the finite overlap between the donor envelopes in Fig.~\ref{fig2}b and d. To verify the deviation from the Heitler-London limit and whether the phase is robust is this regime where the electrons start to delocalise between the two donors, we have computed the STM image based on a calculated two-electron wavefunction which includes electron interactions. This theoretical image, shown in Fig.~\ref{fig2bis}a, is computed using state-of-the-art atomistic modelling capabilities summarised here and detailed in the Methods section. First, starting from the donor pinpointed positions, the one-electron states are obtained by atomistic tight-binding (TB) modelling using a multi-million atom grid. The two-electron states are then obtained using a full configuration interaction method (FCI)~\cite{Tankasala2018, Wang2016} based on $1e$-molecular orbitals. Interface, reservoir and electric field effects are taken into account throughout this modelling framework to ensure an accurate description of the coupled donors spectrum. We found a two electron ground state composed at 67\% of the bonding $A_1{-}A_1$ ground state with an exchange energy of 1.5\,meV (see Supplementary Note 1). As a reference, the Heitler-London limit results in an equal 50\% contribution of both $A_1{-}A_1$ bonding and anti-bonding molecular orbitals since the tunnel coupling vanishes. The theoretical STM image is then obtained by computing the QPWF including STM transport and tip orbital effect~\cite{Usman2016} and is shown in Fig.~\ref{fig2bis}a. We have applied the same filtering procedure to obtain the FFT of this image, shown in Fig.~\ref{fig2bis}b. It shows very similar features compared to the experimental FFT, with notably the presence of valley components at $\pm0.81k_0$ and the diagonal slices which indicate the interference. The resulting $x$ and $y$-valley images are shown in Fig.~\ref{fig2bis}c and d respectively. The valley phases are also pinned to the donor ions and the valley phase differences can be obtained using an identical fitting procedure, giving $\Delta\phi^{FCI}_x=(0.025\pm0.001)^*2\pi$ and $\Delta\phi^{FCI}_y=(0.430\pm0.001)^*2\pi$, in good agreement with the experimental data as the fitted images in Fig.~\ref{fig2bis}e and f show.

We present in Fig.~\ref{fig2bis2}a the results obtained using the same protocol performed on another pair. The two donors are separated by $13a_0\sqrt{2}/2$ along $[110]$ and $9.25a_0\sqrt{2}$ along $[1\bar{1}0]$, i.e. a larger inter-donor distance and a different orientation than the first pair. Its FFT (Fig.~\ref{fig2bis2}b) presents the same characteristics as for pair \#1, with a clear valleys signal around $0.81k_0$ and the presence of diagonal stripes highlighting the geometric interference between the donors. The real-space images resulting from Fourier filtering around the valley signals and their respective fit are shown in Fig.~\ref{fig2bis2}c-f. We note the presence of phase distortions in the $x$-valley image which are different from a succession of phase slips expected for a destructive valley interference pattern. These isolated features are attributed to an instability of the tunnelling tip during the measurement (See Supplementary Note 1), and, importantly, they do not perturb the region of interest for the valley interference and the exchange interaction, i.e. the central region between the two donors. A theoretical STM image was also obtained and analysed, based on a Heitler-London calculation of the two-electron wavefunction as this pair shown an inter-donor distance greater than 7\,nm and therefore falls in this regime. We summarise the phase differences obtained for each pair in Fig.~\ref{fig2sum}, both experimental and theoretical, to notably highlight the matching of the fitted valley phases to the geometrical phase difference expected from the exact lattice pinpointing of each donor atom. Full details on the filtering procedure, as well as on the robustness of the extracted valley phase differences against the dimension of the Fourier space filter can be found in the Supplementary Note 1.

\begin{figure}[htp]
\begin{center}
\includegraphics[width=8.8cm]{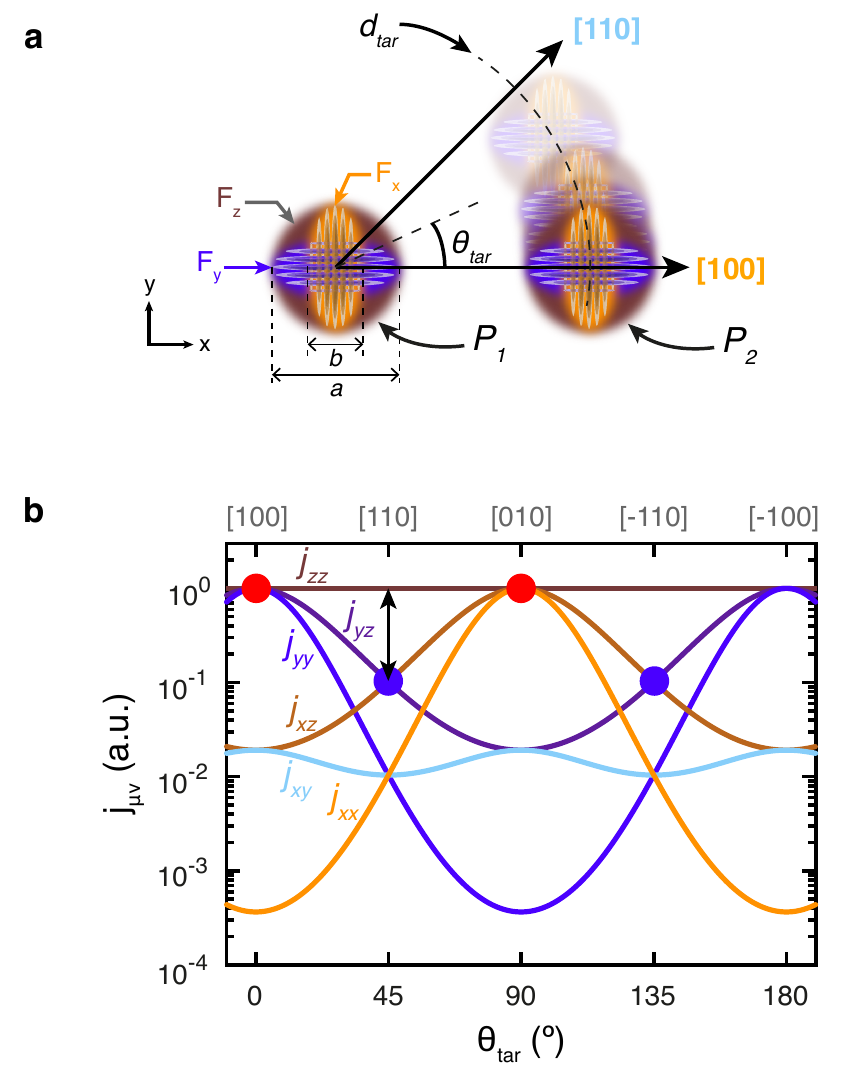}
\caption{ \textbf{Donor envelope overlap and exchange interaction.} \textbf{a} Two donors placed in the same $xy$-plane at a target distance ${d_{tar}}$ and angle $\rm{\theta_{tar}}$ defined from [100]. The exchange interaction can be considered as a sum of valley interfering envelope overlaps $j_{\mu\nu}$ between the $\pm x$ (orange), $\pm y$ (purple) and $\pm z$-valleys (brown). The mass anisotropy results in each envelope orbital, $F_x$, $F_y$ or $F_z$, to be constricted along its own direction. For instance $F_x$ has a small envelope radius $b$ along $x$, and a large envelope radius $a$ along $y$ and $z$. \textbf{b} Normalised envelope weights $j_{\mu\nu}$ plotted vs. $\theta_{tar}$ for $d_{tar}{=}12$\,nm, using $b/a{=}0.52$ and $a{=}2.8$\,nm. Because of orbital anisotropy, the $j_{yy}$,  $j_{yz}$, and $j_{zz}$ terms (respectively $j_{xx}$,  $j_{xz}$, and $j_{zz}$ terms) are degenerate around [100] (respectively [010], red spots), while $j_{zz}$ dominates around [110] and $\rm{[-110]}$ (blue spots).}
\label{fig4a}
\end{center}
\end{figure}

The fitting procedure we have developed here also allows to investigate the envelope part of the donor valley states, whose characteristic extent are represented by the red dashed lines in Fig.2-4, for each donor and each valley. Contrarily to a simple $1s$-orbital with a spherical symmetry, the envelopes are here clearly ellipsoidal. This feature originates from the silicon mass anisotropy, which results in each valley orbital to present a small envelop radius $b$ along their longitudinal direction and a large radius $a$ along the two transverse directions~\cite{Kohn1955, Saraiva2016} (see Fig.~\ref{fig4a}a). The values $b/a$ obtained across all experimental and theoretical image fits average to 0.52 which is in good agreement with single donors measurements~\cite{Saraiva2016} (see Supplementary Note 1) and with other predictions~\cite{Koiller2001,Pica2014,Gamble2015}. Our experimental results demonstrate the existence of a valley interference effect between donor pairs which present a finite overlap between their wavefunctions. It is noteworthy that the tunnel current probes the extent of the wavefunction at the silicon/vacuum interface. In our case, this means that the tunneling tip probes the overlap between the tails of sub-surface donor wavefunctions, which is precisely the essence of what the tunnel and exchange interactions rely upon.

\subsection{Exchange variations analysis for atomically precise donor qubit devices}
Valley interference between neighbouring donors impact the exchange interaction in a non-trivial manner. To detail this effect we have developed a model based on an effective mass Heitler-London formalism~\cite{Koiller2001}. Importantly, this phenomenological effective mass model (P-EM) strongly relates to the valley phase differences $\Delta\phi_{\mu}$ and to the ratio $b/a$ of the donor envelope which have been experimentally investigated above. The P-EM model decomposes the exchange interaction into a sum of oscillating terms, weighted according to 6 possible envelope terms $j_{\mu\nu}$, as follows (see Supplementary Note 3):

\begin{equation}
J(\vec{R})=\sum_{\substack{\mu,\nu= \\ \pm \{x, y, z\}}} j_{\mu\nu}(\vec{R},a,b)\cos(\Delta\phi_{\mu}(\vec{R})\pm \Delta\phi_{\nu}(\vec{R}))
\label{exc}
\end{equation}

\noindent where $\mu$ (respectively $\nu$) denotes the valley in which the first (respectively second) electron is exchanged between the two donors. Qualitatively, each weight $j_{\mu\nu}$ is a four term-product, between two envelope orbitals $F_{\mu}$ separated by $\vec{R}$, and two envelope orbitals $F_{\nu}$, likewise also separated by $\vec{R}$. The $1s$-like nature of the donor orbitals causes these envelope terms $j_{\mu\nu}$ to vary exponentially with the inter-donor distance, with a characteristic length scale related to the major envelope radius $a$. As $k_\mu$ is close to $k_0$, the valley phase differences evolve rapidly with site to site changes in the lattice position of the donors, which can strongly modulate the exchange. Both envelope and valley dependencies hence suggest that the exchange coupling will change significantly by the displacements of dopants, hence requiring atomic scale dopant placement accuracy~\cite{Koiller2001,Song2016} for stability. To date the most precise dopant placement technique is obtained using STM lithography~\cite{Fuechsle2012}, with the fabrication of in-plane atomic devices where donors can be placed within $\pm 1$ lattice site precision. In the following, we investigate the exchange variations which can be expected using this unique device fabrication capability.

\begin{figure*}[htp]
\begin{center}
\includegraphics[width=18cm]{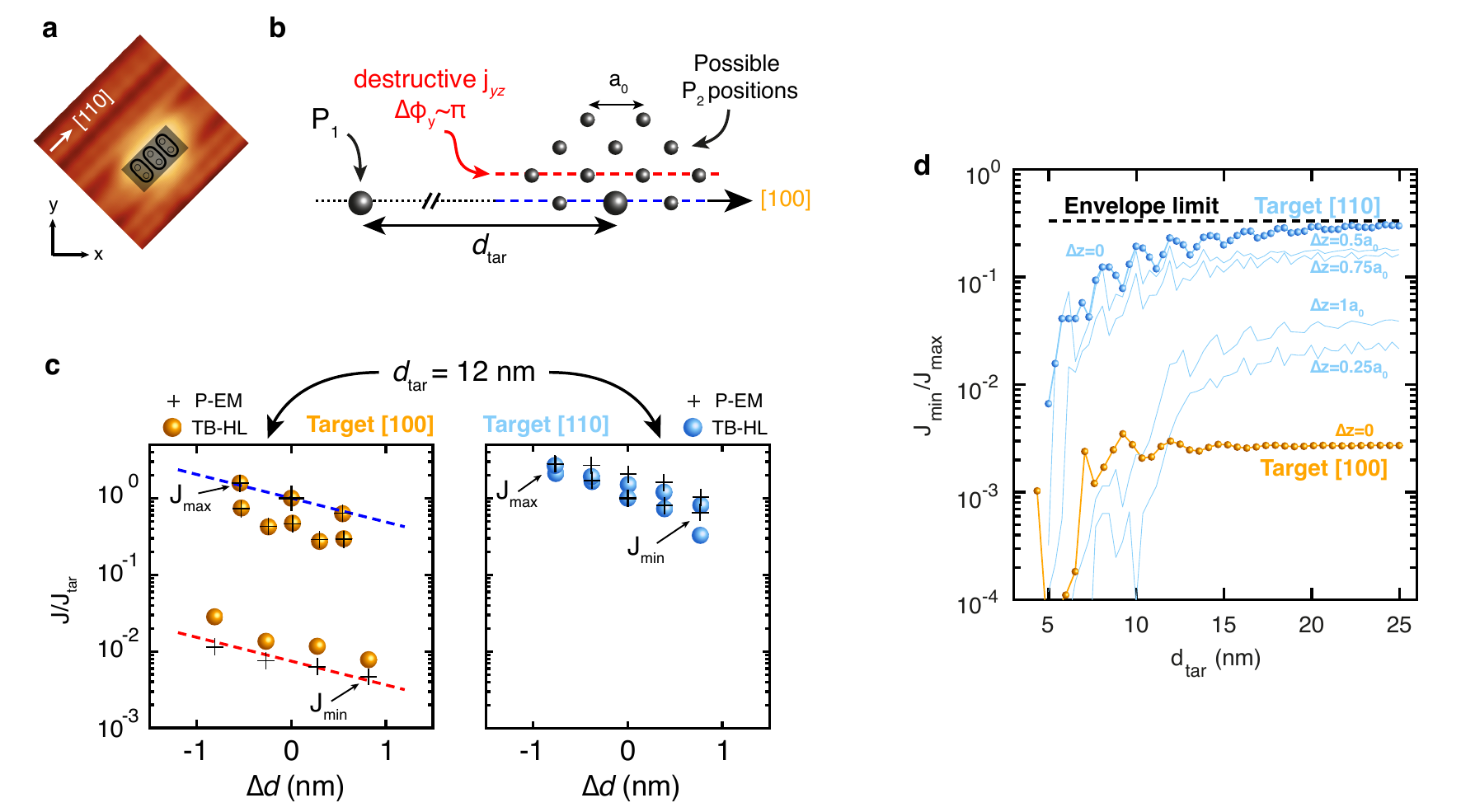}
\caption{ \textbf{Limited exchange in-plane variations for STM placed donors.} \textbf{a} Single donors can be incorporated in a patch of 3 desorbed hydrogen dimers on the silicon surface, giving 6 possible positions for each donor. \textbf{b} There are 12 non-equivalent position configurations for $P_2$ for a target along [100], taking the convention to fix $P_1$ at the origin. A target along [110] results in 10 possible configurations. \textbf{c} Left: plot of the 12 possible exchange values along [100], normalised over the target one, plotted vs the distance difference $\Delta d=d-d_{tar}$ with $d_{tar}=12$\,nm, for both the P-EM (black crosses, the large one denotes the target) and TB-HL (orange balls) models. The configurations along the red dashed line (same as \textbf{b}) result in destructive $j_{yz}$ terms and their exchange values are reduced by more than two orders of magnitude compared to the configurations perfectly aligned with [100] (blue dashed lines). Right: the exchange variations for a target along [110] are limited to less than a factor of 10 as $j_{zz}$ dominates, making the exchange interaction insensitive to in-plane valley interference. \textbf{d} Ratios $J_{min}/J_{max}$ vs $d_{tar}$ between 5 and 25\,nm for both [100] (orange) and [110] (blue). The impact of in-plane valley interference vanishes for target distances beyond 12\,nm along [110], to bound exchange variations to the envelope limit. Destructive $y$-valley positions, and the degeneracy between $j_{yz}$, $j_{yy}$ and $j_{zz}$, are always present along [100] independently of the target distance, resulting in large exchange variations. The light blue lines show the exchange variations if the two donors are in different atomic $z$-planes, resulting in a finite $z$-valley phase difference.} 
\label{fig4b}
\end{center}
\end{figure*}

The relative weights of the 6 different envelope terms $j_{\mu\nu}$ are plotted in Fig.~\ref{fig4a}b for a target distance $d_{tar}$ of 12\,nm as a function of the in-plane angle $\theta_{tar}$ defined with respect to [100]. To parametrise the P-EM model, the valley momentum and anisotropy ratio are set to the experimentally obtained average values, i.e. $k_{\mu}{=}0.81k_0$ and $b/a{=}0.52$. The large envelope radius $a$ is set to 2.8\,nm obtained from fitting Heitler-London calculations along [110] (See Supplementary Note 3). The weights ($j_{xx}$, $j_{xy}$, $j_{xz}$, $j_{yy}$, $j_{yz}$ and $j_{zz}$) evolve smoothly with $\theta_{tar}$ (and $d_{tar}$), since they do not contain any valley interference, but their ratios are not constant and vary by several orders of magnitude. This is a direct consequence of the anisotropic and exponential nature of the donor envelope. Along [100], at $\theta_{tar}=0^\circ$, the $j_{yz}$ terms are degenerate with $j_{zz}$ and $j_{yy}$, and largely dominate over any term where an electron is exchanged in a $x$-valley. This can be understood as mass anisotropy results in $F_{x}$ to have a minor envelope radius $b$ along ${x}$, hence the product between two $F_{x}$ orbitals separated along [100] is smaller than its $F_{y}$  and $F_{z}$-orbital counterparts, as they both have a major envelope radius $a$ along $x$ instead. The degeneracy between $j_{yz}$, $j_{zz}$ and $j_{yy}$ arises from symmetry as the products between $y$ and $z$ orbitals are equal along [100].

The envelope weight anisotropy influences the way valley interference impact the exchange coupling. In order to get further insight on its role, we must consider the precise placement scheme provided by STM lithography. Each donor is stochastically incorporated within $\pm 1$ lattice site precision in a patch of 3 desorbed hydrogen dimers (see Fig.~\ref{fig4b}a). This placement precision results along [100] in 12 non-equivalent configurations for the donors separation shown in Fig.~\ref{fig4b}b, taking the convention to fix $\rm{P_1}$ at the origin, (see Supplementary Note 3). From pair to pair, different configurations will be stochastically obtained, resulting in the phase terms $\Delta\phi_x$ and $\Delta\phi_y$ to vary accordingly, and to potentially become destructive. Along [100], the degeneracy and dominance of the $j_{yy}$, $j_{zz}$ and $j_{yz}$ terms implies that the sum expressed in Eq.~\ref{exc} can be reduced to these terms, which only involve $\Delta\phi_y$ and $\Delta\phi_z$. Hence, $\Delta\phi_x$ does not influence the exchange and can be ignored along [100]. Furthermore, $\Delta\phi_z{=}0$ since the $z$-valleys are constructive anywhere in the $xy$-plane, and only $\Delta\phi_y$ matters along [100]. The 12 values of exchange obtained from these configurations are plotted in Fig.~\ref{fig4b}c for a target distance of 12\,nm. Assigning each of them to their respective position configuration allows to understand the observed spread. The values can be separated in 4 different groups according to their respective $\Delta\phi_y$ since it is the only phase term of interest in this case. Among each group, the exchange values evolve monotonically since their valley phase terms are the same, leaving only an envelope dependence. The configurations perfectly aligned with [100] (blue dashed line in Fig.~\ref{fig4b}b-c) lead to constructive $y$-valleys, i.e. $\Delta\phi_y=0$ and a maximised exchange coupling. The fast damping of exchange oscillations purely along [100] was already pointed out in previous work~\cite{Koiller2001,Pica2014,Gamble2015}, which can result in favouring [100] as the target axis if dopant placement accuracy is not appropriately considered. In fact, the configurations misaligned with [100] result in finite $\Delta\phi_y$ and hence to a reduced exchange energy. In particular, in-plane positions off the [100] axis by $a_0/2$, i.e. the closest positions to [100] (red dashed line in Fig.~\ref{fig4b}b-c), result in destructive $y$-valley interference and hence negative $j_{yz}$ contributions to the exchange. These negative contributions almost perfectly cancel out the constructive $j_{yy}$ and $j_{zz}$ terms in the exchange equation, and the exchange energy is reduced by more than two orders of magnitude for these specific positions~\cite{Koiller2001}. It is important to note that these destructive configurations are common to any previous theoretical work, although they had different conclusions because of different approaches in dopant placement accuracy (see Supplementary Note 3). Moreover, they are always present for any inter-donor distance since they result from the envelope weight degeneracy between $j_{yy}$, $j_{yz}$ and $j_{zz}$ inherent to the [100] axis. These two arguments are crucial to definitely rule out [100] as a favourable placement axis.

Considering placement accuracy is essential when looking to fully understand the role of valley interference and envelope anisotropy on the exchange interaction. Some approaches have considered random dopant placement around a target~\cite{Koiller2001,Testolin2007,Gamble2015,Song2016}, but the actual in-plane placement accuracy at any in-plane angle $\theta_{tar}$ that STM lithography can ultimately offer has been overlooked. Away from [100] at finite $\theta_{tar}$, the envelope anisotropy breaks the degeneracy between the $j_{yy}$, $j_{zz}$ and $j_{yz}$ terms (see Fig.~\ref{fig4a}b). The $j_{zz}$ term becomes dominant over any term involving an electron exchanged in a $F_x$ or $F_y$ orbital, as the $F_z$ orbitals are the only ones to face each other across their major envelope radius $a$ independently of $\theta_{tar}$ (as their minor envelope radius is out of the plane). The prevalence of $j_{zz}$ is maximised at $\theta_{tar}=45^\circ$, i.e. for donors separated along [110]. There, a different symmetry condition is reached with a degeneracy occurring between the $F_{x}$  and $F_{y}$  orbital products since $x{=}y$ along this direction. The following order is hence obtained, with the $j_{zz}$ terms dominating over the degenerate $j_{xz}$ and $j_{yz}$ terms, themselves dominating over the $j_{xx}$, $j_{xy}$ and $j_{yy}$ terms, which are also degenerate by symmetry. Placing donors along the [110] orientation results in finite $\Delta\phi_x$ or $\Delta\phi_y$, which can vary according to the configuration specifically obtained from pair to pair, and hence to destructive $j_{xz}$ or $j_{yz}$ terms. However, they both have a negligible impact on the value of exchange since $j_{zz}$ dominates. As a result, the possible exchange values obtained for two donors aimed along the [110] axis are much more constrained and vary by less than an order of magnitude as shown in Fig.~\ref{fig4b}c. To summarise these results we plot in Fig.~\ref{fig4b}d the ratio $J_{min}/J_{max}$ defined for each target distance and orientation. Along [100], the exchange energy presents large variations independent of target distance as the degeneracy between $j_{yy}$, $j_{yz}$ and $j_{zz}$ remains. However, along [110] the in-plane valley interference $\Delta\phi_x$ or $\Delta\phi_y$ do not impact the exchange coupling for target distances beyond 12\,nm as the variations reach an asymptotic limit set by envelope considerations only (see Supplementary Note 3). As for the [100] case, this result is in agreement with previous work~\cite{Koiller2001,Pica2014,Gamble2015} (See supplementary Note 3). It is revealed here through an atomic-scale understanding of the interplay between envelope anisotropy, degeneracies, valley interference and dopant placement accuracy. Furthermore, the dominance of the $j_{zz}$ envelope terms along [110] results in $\Delta\phi_z$ to be the only relevant valley phase difference, and thus to the exchange variations to be arranged as a function of the depth difference, i.e. atomic planes, between the donors. The resulting variations with respect to the maximum in-plane exchange value are shown in Fig.~\ref{fig4b}d for depth variations included within $\pm$1\,$a_0$. Only the ${a_0/4}$ and $1a_0$ planes show a significant reduction of the exchange interaction. Moreover, for target distances beyond 10\,nm none of these planes lead to exchange variations as large as for the in-plane exchange variations obtained around the [100] direction, which are the largest within $\pm$1\,$a_0$ variation in $z$-placement for this direction (See Supplementary Note 3).

We also compared the variations obtained from our P-EM model to Heitler-London calculations based on the tight-binding wavefunctions (TB-HL), for which we already demonstrated in Fig~\ref{fig2} an excellent agreement with the experiment at the wavefunction level. The TB-HL exchange calculations shown in Fig.~\ref{fig4b}c confirm two orders of magnitude of exchange variations for donors placed along the [100] direction because of the presence of destructive $y$-valleys configurations. They also confirm that exchange variations can be reduced to less than an order of magnitude along [110]. A more detailed level of comparison between TB and P-EM models is discussed in the Supplementary Note 3. Hence, TB formalism accurately models the donor wavefunction's details and can be further used to predict the properties of advanced donor-based quantum devices which notably include electric fields. Finally, our understanding of the interplay between valley interference, envelope anisotropy and atomic placement on the exchange interaction which we developed from our P-EM model can be tied back to the original pair which has been experimentally measured. In order to obtain a normalisation constant for the exchange energies, we have fitted the exchange energy obtained from tight-binding calculations in the Heitler-London limit along [110] for the bulk case (See Supplementary Note 3). We notably obtain excellent agreement on the value of the valley momentum $k_{\mu}$ and on the anisotropy $b/a$ with the experimental ones. Equipped with this calibration, we can predict an effective mass exchange value for pair \#1, which we found to be 1.50\,meV, in excellent agreement with the FCI value calculated and mentioned above. Furthermore, we have also computed a nearby case of pair \#1, where donor $P_1$ was brought from the $z{=}5.5a_0$ to the $z{=}6.5a_0$ atomic plane. This shift results for these two donors to have the same $x$ and $y$ coordinates but to be separated by $\Delta z{=}0.25a_0$, which is a very destructive plane difference for the $z$-valleys as seen in Fig.~\ref{fig4b}d for such pair in the neighbourhood of [110]. Indeed, the P-EM model yields an exchange of 0.133\,meV and the FCI calculations 0.108\,meV, again in excellent agreement with each other. We note that for such inter-donor distance the $z$-arrangement of the exchange values along [110] is not fully formed yet, i.e. $\Delta\phi_x$ and $\Delta\phi_y$ are non-negligible for these two cases, nevertheless this quantitative agreement clearly demonstrates the influence of the valley interference on the exchange interaction, as both values for pair \#1 and its nearby cases are much lower than the maximum exchange values of the order of 10\,meV that can be obtained in this neighbourhood.

 \section{Discussion}

The direct measurement and quantification of valley interference between donor states at the atomic-scale using STM which was realised here provides a detailed level of understanding of their impact on exchange variations~\cite{Koiller2001,Gamble2015,Song2016}. Importantly, we found that the exchange interaction along the [110] direction is dominated by the $j_{zz}$ term and hence insensitive to in-plane valley interference $\Delta\phi_x$ and $\Delta\phi_y$. The agreement between atomistic calculations and the P-EM model which relies upon parameters which we have here assessed experimentally, further establishes the importance of the envelope radius anisotropy and of valley interference to fully understand the behaviour of the exchange interaction at the atomic scale. Using the current state of the art STM lithography~\cite{Fuechsle2012, Weber2014, Koch2019} with $\pm$\,1 lattice site precision, the preferential crystallographic placement along [110] can be leveraged together with an in-plane placement resulting in $\Delta\phi_z=0$. As a result, STM lithography enables to engineer donor devices where the exchange can be totally immune to valley interference, hence achieving a true semiconductor vacuum where complex degeneracies of the band structure can be ignored for coupled donors. The exchange variations are minimised to a factor of less than 10 between configurations where dopants are moved by $\pm$1 lattice site. This factor is only due to change in the envelope overlap between the wavefunctions as it is the case for vacuum systems. Importantly, by symmetry, and as confirmed by our results in Fig.~\ref{fig4a}b, the exchange coupling along the [-110] direction (perpendicular to [110]) is also dominated by $j_{zz}$ and therefore also protected from in-plane valley interference. As such it is possible to place donors using STM lithography along both the [110] and [-110] directions to create 1D and 2D arrays with reduced exchange coupling variations between nearest neighbours. In our work, the donors were found at different atomic planes which we attribute to an annealing step at $600\,^{\circ}$C performed during sample fabrication in order to flatten the surface for tunneling spectroscopy purposes. Experimental progress has been made to minimise the segregation of highly doped phosphorus monolayers~\cite{Keizer2015,Wang2018a}, which should be further reduced for single donors as segregation and diffusion constants strongly depends on dopant concentration~\cite{Hu1983}. Single donor segregation mechanisms are predicted to activate from $250\,^{\circ}$C~\cite{Bennett2009}, which bulk donor qubit device fabrication can withstand~\cite{He2019}. These results motivate our scheme to keep the donors in the same plane around the [110] axis to maximise the exchange interaction uniformity.

Controllable exchange coupling between atom qubits is a key requirement for two-qubits gates, which must be performed with high fidelities and high speed in views of fault-tolerant quantum computing architectures~\cite{Fowler2012, Hill2015}. Exchange variations from pair-to-pair can create rotation errors as the CNOT gate length is calibrated to a particular target value, which can impair the operation of a quantum processor. To avoid these errors, a composite CNOT gate can be performed with a so-called BB1 sequence instead of a single pulse~\cite{Hill2005,Testolin2007}. This composite CNOT gate can be decomposed in a set of single and two-qubit rotations, with the advantage to maintain a high-fidelity above 99.9\%, i.e. above quantum error correction thresholds, for any qubit pair despite a 10\% error in the characterisation of the exchange coupling. For donors separated by 12\,nm along the [110] direction, our TB-HL calculations give exchange values ranging from 11 to 93\,MHz. These values can be used together with known single qubit gate times (340\,ns~\cite{Pla2012}) to obtain CNOT gate times varying by less than 1\% across an array (See Supplementary Note 4), with an average value of $1.2\mu s$, for fully characterised exchange values. This average CNOT gate time is dominated by the single qubit rotation time for this target distance, and the spread could be overcome by tuning the exchange values with each other electrically~\cite{Wang2016}. Such tuning would be impossible for donors placed along [100] because of too large exchange variations. The operation time expected for the out-of-plane configurations within one monolayer do not exceed $1.34\mu s$. For exchange values characterised to 90\%, adding the correcting sequence to maintain the fidelity would result in CNOT operation times average to $2.3\mu s$ with a spread limited to 4\% for the in-plane configurations, and a maximum time of $3.7\mu s$ for out-of-plane configurations with only 6 out of 82 total configurations exceeding $3\mu s$. Whilst these operation times are slower compared to other spin-based CNOT gates~\cite{Veldhorst2015, Watson2018, Zajac2017}, they are well below donor coherence times which can approach seconds using dynamical decoupling sequences~\cite{Tyryshkin2011}.\\

As already mentioned, exchange and tunnel interactions are closely related since they both rely on wavefunction overlap considerations. From our exchange analysis, we obtain tunnel coupling values varying by less than a factor 5 for inter-donor distances down along [110] to 5\,nm (see Supplementary Note 3). This distance is the lower bound at which the Heitler-London description is valid, as tunnel coupling values are there predicted to exceed 10\,meV, i.e. becoming a substantial fraction of the 47\,meV charging energy. Expected tunnel coupling values~\cite{Gamble2015} up to target distances of 12\,nm along [110] remain above 200$\mu$eV for in-plane configurations, and above 60$\mu$eV for out-of-plane ones, all much larger than the achievable electronic temperature of about 50\,mK (i.e. about 5$\mu$eV). This opens the way to the study of Fermi-Hubbard problems robust to this level of disorder, as for the case of dimerised chains~\cite{Su1979,Le2019}, with the competition between topological gaps originating from the tight-binding picture~\cite{Slot2017, Drost2017, Ni2019} and on-site interactions. The prospect to extend studies to the regime of strong tunneling interactions and to 2D systems makes donors in silicon a standout platform for quantum simulation of the Fermi-Hubbard model.\\

In summary, starting from direct real-space measurements of donor wavefunctions, we have been able to unambiguously quantify valley interference between donor atoms in silicon in real space and validate the predictions of existing theories. Driven by this experimental approach, we consider the dopant placement precision offered by STM lithography for a detailed understanding of the interplay between valley interference, envelope anisotropy on the exchange interaction. In full agreement with previous theoretical work, our results identify the [110] crystallographic direction as optimal for building 2D donor arrays where we predict less than an order of magnitude variation in exchange and tunnel couplings. We envision this fabrication strategy, in conjunction with quantum control schemes~\cite{Testolin2007} and exchange tuning mechanisms~\cite{Wang2016}, to be a key component in leveraging the exceptional coherence of donor qubits in silicon towards scalable quantum simulators and quantum processors.

\section*{Methods}
\subsection{Sample preparation}

Samples were prepared in ultrahigh vacuum (UHV) with a pressure lower than $10^{-10}$\,mbar, starting from a commercial n-type As doped wafer with a resistivity of $\rm{0.001-0.003\,\Omega.cm}$. Samples are first flash annealed three times around $\rm{1150\,^{\circ}C}$ for a total of 30\,s. After the final flash anneal, the temperature was rapidly quenched to $\rm{800\,^{\circ}C}$, followed by slow ($\rm{1\,^{\circ}C/s}$) cooling to obtain a flat $\rm{2\times1}$ surface reconstruction. Under these conditions, a layer of $\rm{\sim15\,nm}$ from the Si surface is depleted from As dopants. P dopants are incorporated at this stage in Si by submonolayer phosphine ($\rm{PH_3}$) dosing with a sheet density of $\rm{5\cdot10^{11}\,cm^{-2}}$. This low-dose P $\rm{\delta}$-layer was overgrown epitaxially by $\rm{\sim2.5\,nm}$ of Si. Growth parameters such as temperature and flux were chosen to achieve minimal segregation and diffusion whilst preserving a flat surface for STM imaging and spectroscopy purposes. Notably, the first nanometer is a lock-in layer grown at room temperature~\cite{Keizer2015}. Subsequent growth alternates between $\rm{250\,^{\circ}C}$ and $\rm{450\,^{\circ}C}$ with a duration ratio of 3/1. A $\rm{600\,^{\circ}C}$ flash follows for 10\,s to flatten the surface. The surface is finally hydrogen passivated at $\rm{340\,^{\circ}C}$ for 10\,min under a flux of atomic H produced by a thermal cracker, in a chamber with a $\rm{10^{-7}\,mbar}$ pressure of molecular hydrogen. STM measurements are taken in the single-electron transport regime described in ref.~\cite{Salfi2014, Voisin2015}.

\subsection{Measurement techniques}

The electrical measurements were carried out at 4.2\,K in a scanning tunnelling microscope (Omicron LT-STM). Both sample fabrication and measurements are done in ultrahigh vacuum with a pressure lower than $10^{-10}$\,mbar. The tunnel current $I$ was measured as a function of the bias voltage $U$ using ultralow noise electronics including a transimpedance amplifier. The differential conductance d$I$/d$U$ shown in Supplementary Note 2 was obtained by numerical differentiation. Spatially resolved measurements of donor pairs quantum state were acquired using the multi-line scan technique, where the topography is recorded at ${U= -1.45\,V}$ during the first pass, and played during the second pass in open-loop mode with the current $I$ recorded at the bias mentioned in the caption of the corresponding figures. The sample fabrication described above results in the donor pairs to be measured in the sequential transport regime, with a first tunnel barrier with tunnel rate $\rm{\Gamma_{in}}$ occurring from the highly doped substrate annealing, and the second tunnel barrier with tunnel rate $\rm{\Gamma_{out}}$ being a combination of the Si overgrowth after P deposition and the vacuum barrier, mainly dominated by the latter and tip-sample distance. Additional information regarding STM images and spectroscopy analysis can be found in the Supplementary Note 1.

\subsection{Atomistic simulations}

Single-particle energies and wave functions for P donor in silicon are computed by solving a $sp^3d^5s^*$ tight-binding Hamiltonian~\cite{Boykin2004a}, where the P atom is represented by central-cell corrections including donor potential screened by non-static dielectric function~\cite{Usman2015}, and the P-Si nearest-neighbour bond-lengths are modified in accordance with the published DFT prediction~\cite{Overhof2004}. The size of simulation domain (Si box) consists of roughly four million atoms with closed boundary conditions in all three spatial directions. The effect of surface strain due to $\rm{2\times1}$ surface reconstruction is included in the tight-binding Hamiltonian by properly displacing surface Si atoms and by modifying the inter-atomic interaction energies in the tight-binding Hamiltonian~\cite{Usman2016}. The calculation of STM images of donor wave functions follows the published methodology~\cite{Usman2016}, where tight-binding wave function is coupled with Bardeen's tunneling theory~\cite{Bardeen1961} and derivative rule of Chen~\cite{Chen1990}. In our STM measurements, dominant contribution is from $d_{z^2}$ orbital in STM tip. For two-particle STM images, two-electron wave functions are computed from full configuration interaction approach~\cite{Tankasala2018}. The STM image represents a quasi-particle wave function resulting from 2e to 1e transition~\cite{Rontani2005}. The resulting quasi-particle state is used to compute tunnelling matrix element described in the Supplementary Note 1. The exchange calculations shown in Fig.~\ref{fig4b} are computed either from the corresponding atomistic tight-binding single-particle wave functions~\cite{Wellard2003}, or from a phenomenological effective mass model detailed in the Supplementary Note 3, both based on Heitler-London formalism.

\section{Acknowledgements}
We thank J. Keizer for advice on sample fabrication, C.Hill, M.J. Calderon and S.N. Coppersmith for valuable discussions, and J.K. Gamble \textit{et al.} for publishing a complete data set of tunnel coupling calculations~\cite{Gamble2015}. We acknowledge support from the ARC Centre of Excellence for Quantum Computation and Communication Technology (CE170100012), Silicon Quantum Computing Pty Ltd., and from the U.S. Army Research Office (W911NF-08-1-0527). J.S. acknowledges support from an ARC DECRA fellowship (DE160101490). The authors acknowledge the use of computational resources from NanoHUB, the Network for Computational Nanotechnology at Purdue University, and from the Australian Government through the Pawsey Supercomputing Centre under the National Computational Merit Allocation Scheme (NCMAS). This work used the Extreme Science and Engineering Discovery Environment (XSEDE) ECS150001, which is supported by National Science Foundation Grant No. ACI-1548562~\cite{Towns2014}.

\section{Author contributions}

S.R. designed the project. J.B. and  B.V. conducted the STM experiments and carried out the STM image analysis, with J.S., M.Y.S. and S.R. providing input. M.U., B.V., A.T., J.S. and J.B. pinpointed the dopant positions. M.U., A.T. and R.R. performed the single donor tight-binding calculations. A.T. and R.R. performed the two-electron FCI exchange calculations and computed QPWFs, with J.S. providing input. M.U. calculated the theoretical STM images and computed Heitler-London exchange calculations using tight-binding wavefunctions, with L.C.L.H. and R.R. providing input. B.V. formulated the phenomenological effective mass model and carried out the exchange variation analysis. All authors contributed to data analysis and conclusions. B.V., J.S., M.Y.S., L.C.L.H. and S.R. wrote the manuscript, with input from all authors.

\section{Data availability}
Any data and code used for the purpose of this article is available upon reasonable request. 

\section{Competing interests}
M.Y.S. is a director of the company Silicon Quantum Computing Pty Ltd.
B.V, J.S. and S.R. have submitted a patent application that describes the use of a stabilised exchange coupling for quantum computing with donors in silicon. Other authors declare no competing interests.

\bibliographystyle{aipnum4-1.bst}
\bibliography{P2_ArXiv}


\clearpage

\begin{widetext}

\begin{center}
\large{\textbf{Valley interference and spin exchange at the atomic scale in silicon\\-\\Supplementary Information}\\}
\vspace{0.5cm}
B. Voisin$^{1*}$, J. Bocquel$^{1*}$, A. Tankasala$^{2*}$, M. Usman$^{3,4*}$, J. Salfi$^{1}$, R. Rahman$^{2,5}$, \\M.Y. Simmons$^{1}$, L.C.L. Hollenberg$^{3}$ and S. Rogge$^{1}$\\

\vspace{1em}

\small{{$^1$\, Centre for Quantum Computation and Communication Technology, School of Physics, The University of New South Wales, Sydney, 2052, NSW, Australia} \\
{$^2$\, Electrical and Computer Engineering Department, Purdue University, West Lafayette, Indiana, USA}\\
{$^3$\, Centre for Quantum Computation and Communication Technology, School of Physics, The University of Melbourne, Parkville, 3010, VIC, Australia}\\
{$^4$\, School of Computing and Information Systems, Melbourne, School of Engineering, The University of Melbourne, Parkville, Victoria 3010, Australia.}\\
{$^5$\, School of Physics, The University of New South Wales, Sydney, 2052, NSW, Australia} \\
{$^{*}$\,These authors contributed equally}}
\end{center}
\setcounter{equation}{0}
\setcounter{figure}{0}
\setcounter{table}{0}
\makeatletter
\renewcommand{\theequation}{S\arabic{equation}}
\renewcommand{\thefigure}{S\arabic{figure}}

\section{Supplementary Note 1 - STM image analysis and comparison to TB-FCI theory}
\subsection{STM image correction and filtering}

The STM image of the $2e$ to $1e$ transition is taken following the same procedure as in ref.~\cite{Salfi2014}: we perform a two-pass scan where, for each line of the scan, the topography is recorded during the first pass at large bias $U=-1.45$\,V (corresponding to imaging the valence band states of the silicon surface), and then played in the second pass with the STM tip feedback loop turned off and the bias plunged in the silicon gap at $U=-0.95$\,V, above the $2e$-state transition of the two-donor system (see section S2).\\

The STM images analysis  starts with a drift correction (Fig.~\ref{figS1}a-b): a 2D-linear transformation is applied to the spatial coordinates in order to match the silicon lattice constant, using the topography image. The distance between the silicon dimers of the $2 \times 1$ reconstructed surface should equal $\rm{a_0\sqrt{2}}$, which is obtained via a 2D matrix transformation of the coordinates. We also apply an exponential correction to the tunnel current to compensate for the tip height variations present during the measurement, corresponding to the topography measured in the first pass.\\

A contour filter is applied to the corrected data, with data outside the yellow ellipse shown in Fig.~\ref{figS1}c put to zero with a smooth transition around this contour to avoid artefact resonances in Fourier space. The 2D discrete Fourier transform, obtained numerically from the corrected and filtered data, is shown in Fig.~\ref{figS1}d. The diagonal slices (blue dashed lines) arise from the Fourier transform of the sum of two similar objects shifted in space: let $f(\vec{r})$ be a given spatial function, like the STM image of a single donor, and $F(\vec{k})$ its Fourier transform. The Fourier transform of $f(\vec{r})+f(\vec{r}-\vec{r_0})$ is then $F(\vec{k}) (1+e^{i\phi(\vec{k})})$ with $\phi(\vec{k})=\vec{k}.\vec{r_0}$. The momentum positions $\vec{k_n}$ such as $\phi(\vec{k_n})=\vec{k_n}.\vec{r_0}=n\pi$ result in the Fourier transform to go to zero, which can also be interpreted as a destructive interference condition. The position of the blue dashed lines which correspond to this condition with $\vec{r_0}$ being the relative donor-donor in-plane coordinates give a geometric hint to visualise the valley interference condition: the $x$-valleys, located at  $\mathrm{k}_{x} \sim 0.81\rm{k_0}$, are in-between two slices and are therefore in-phase. On the contrary, the $y$-valleys around $\mathrm{k}_{y} \sim 0.81\rm{k_0}$ fall onto a blue diagonal slice and are then out-of-phase.\\

\begin{figure}[htp]
\includegraphics{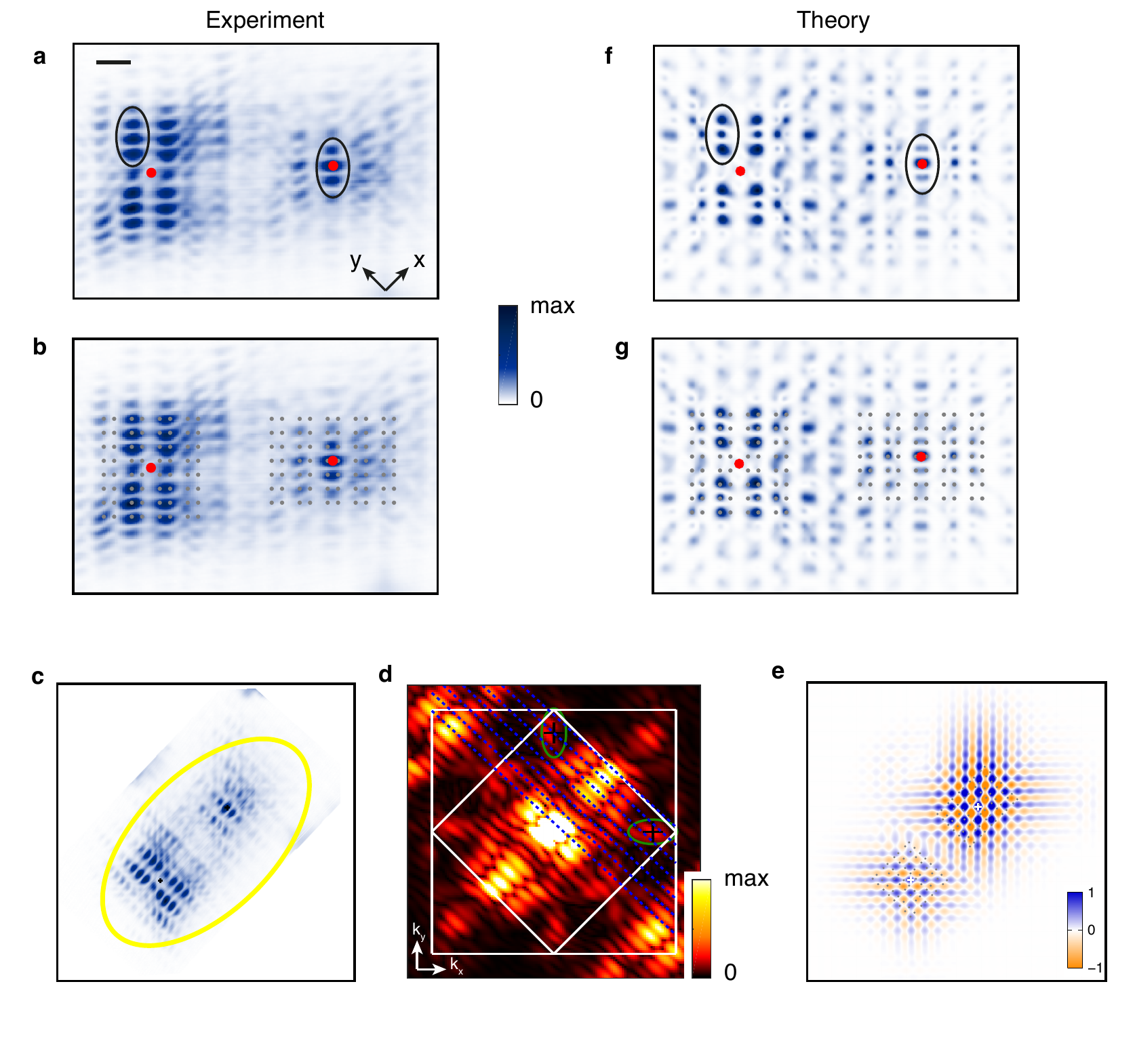}
\caption{ \textbf{Drift correction and valley filtering in Fourier space. Pair \#1. a,} Experimental STM image taken at $\rm{U=-0.95V}$, after drift correction, same as in the main text (pair \#1). Scale bar is 2 nm. The black ellipses indicate examples of features used to pinpoint the exact lattice site position of the phosphorus atoms, according to a procedure developed in ref~\cite{Usman2016}. The red dots indicate the projection of the positions of the P atoms on the surface. \textbf{b,} same as \textbf{a}, with the grey dots indicating the positions of the silicon atoms of the 2$\times$1 reconstructed surface. \textbf{c,} Experimental STM image. The yellow ellipse denotes the contour filter applied to the data before being Fourier transformed. Data is set to zero outside of the ellipse, preserved inside it. A low-passed filter is applied using a 3-point 2D Gaussian kernel to smoothen the transition between the two regions. \textbf{d,} FFT of the STM image (same as main text). The blue dashed lines represent destructive interference conditions, from which the interference condiition of the $x$ and $y$-valley states can be deduced. \textbf{e,} Addition of the $x$ and $y$-valley images shown in Fig.2d of the main manuscript. The grey circles represent the silicon atoms of the 2*1 reconstructed surface. The valley phases, i.e. maxima of the valley signal for each lobe, are pinned to the ion location of each donor, shown by the white crosses. The valley spatial period is clearly different from the lattice spatial periodicities, with also different orientation. \textbf{f-g} Same as \textbf{a-b} for the theoretical STM image computed from TB-FCI molecular state.}
\label{figS1}
\end{figure}
	
In order to quantitatively determine the valley phase difference, the data are multiplied by a Gaussian function in Fourier space (see equations~\ref{maskx} and ~\ref{masky}). These Gaussian functions are centred around $\mathrm{k}_{x} \sim 0.81\rm{k_0}$ or $\mathrm{k}_{y} \sim 0.81\rm{k_0}$ to focus on the $x$ or $y$-valleys, respectively. The filtered Fourier data are transformed back to real space to obtain the images shown in Fig. 2-3-5 of the main text. The variances $f_r$ and $f_t$ were carefully chosen in order to capture the relevant valley signal while avoiding other components of the Fourier transform as done in~\cite{Saraiva2016}, and their impact on the fits are discussed in detail below. The green ellipses in each Fourier image of the main text represent the $2\sigma$-contour of the Gaussian filter masks which were used for both experimental and theoretical images, with $f_r{=}0.1(2\pi/a_0)$ and $f_t/f_r{=}0.7$. Back to real space, the amplitude $A_{\mu}(x,y)$ of each 2D valley image shown in the main text was fitted using the following equations~\ref{fiteqx} and ~\ref{fiteqy} :

\begin{equation}
\begin{gathered}
\mathrm{mask}_x(k_x,k_y)=e^{-\frac{1}{2}[(\frac{(k_x/(2\pi/a_0)-0.81)}{f_r})^2+(\frac{(k_y/(2\pi/a_0)-0.81)}{f_t})^2]}
\label{maskx}
\end{gathered}
\end{equation}

\begin{equation}
\begin{gathered}
\mathrm{mask}_y(k_x,k_y)=e^{-\frac{1}{2}[(\frac{(k_y/(2\pi/a_0)-0.81)}{f_r})^2+(\frac{(k_x/(2\pi/a_0)-0.81)}{f_t})^2]}
\label{masky}
\end{gathered}
\end{equation}

\begin{equation}
\begin{gathered}
\mathrm{Fit1}_x(x,y)=A_1e^{-(\frac{x-x_1}{\sqrt{2}b_1})^2-(\frac{y-y_1}{\sqrt{2}a_1})^2}*e^{-\frac{(x-x_1)^2+(y-y_1)^2}{2a_1^2}}*\cos(\frac{2\pi}{\lambda_{x}}(x-x_1))+\\
A_2e^{-(\frac{x-x_2}{\sqrt{2}b_2})^2-(\frac{y-y_2}{\sqrt{2}a_2})^2}*e^{-\frac{(x-x_2)^2+(y-y_2)^2}{2a_2^2}}*\cos(\frac{2\pi}{\lambda_{x}}(x-x_2))
\label{fiteqx}
\end{gathered}
\end{equation}

\begin{equation}
\begin{gathered}
\mathrm{Fit1}_y(x,y)=A_1e^{-(\frac{x-x_1}{\sqrt{2}a_1})^2-(\frac{y-y_1}{\sqrt{2}b_1})^2}*e^{-\frac{(x-x_1)^2+(y-y_1)^2}{2a_1^2}}*\cos(\frac{2\pi}{\lambda_{y}}(y-y_1))+\\
A_2e^{-(\frac{x-x_2}{\sqrt{2}a_2})^2-(\frac{y-y_2}{\sqrt{2}b_2})^2}*e^{-\frac{(x-x_2)^2+(y-y_2)^2}{2a_2^2}}*\cos(\frac{2\pi}{\lambda_{y}}(y-y_2))
\label{fiteqy}
\end{gathered}
\end{equation}

which gives access to the $(x,y)$-coordinates of the absolute maxima and to the anisotropy $(b/a)$ for each donor and each valley, as well as a common valley wavelength $\lambda_{\mu}$. The phase differences are obtained using $\Delta\phi_x=|x_2-x_1|/\lambda_{x}$ and $\Delta\phi_y=|y_2-y_1|/\lambda_{y}$, modulo $2\pi$, and their corresponding confidence intervals are derived from the fit parameters deviations. Fig.~\ref{figS1}e shows that the maxima of the sum of the $x$ and $y$-valley image fall on the ion positions, which means that the valley phase is pinned to the ion for each donor.

\begin{figure}[htp]
\hspace{-0.05cm}
\includegraphics{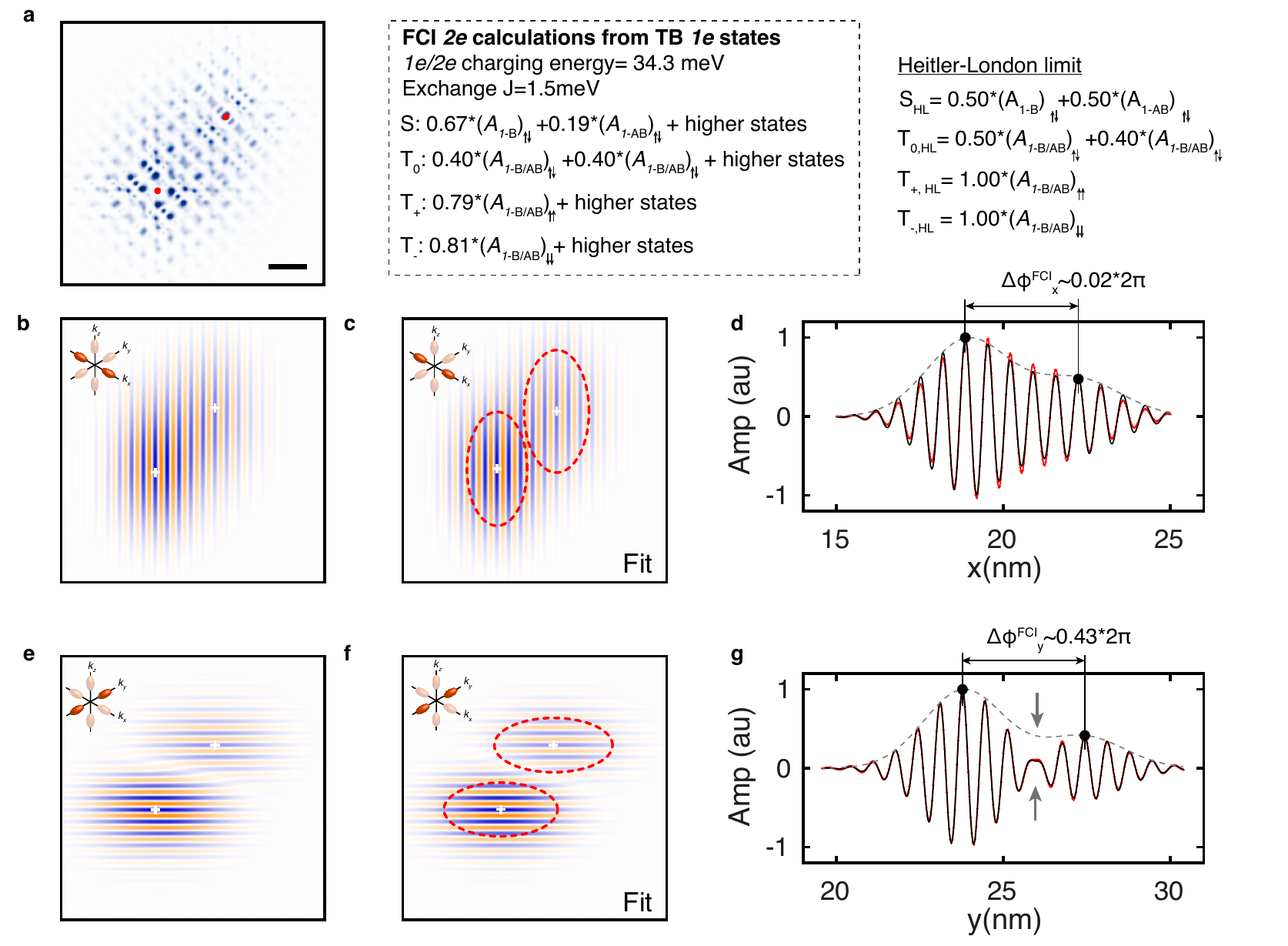}
\caption{\textbf{Details on FCI calculations and theoretical valley phase difference for pair \#1. a,} Theoretical STM image (scale bar is 2\,nm) based on $2e$-FCI calculations, same as main text. Details are given about the charging energy, exchange energy and about the molecular orbital composition of the $2e$ singlet and triplet states. The comparison to a pure Heitler-London state is also given on the right-hand side. \textbf{b,} Theoretical STM image filtered around the $x$-valleys as shown in the main text. \textbf{c,} Fitted valley image using eq.~\ref{fiteqx}, same as main text. \textbf{d,} Line cut taken across the two ions for both images. In-between the two donors, the amplitude of the valley signal remains at the maximum of the sum of the envelope of both donors (grey dashed line) as the $x$-valley interference are constructive. \textbf{e-g} Same as \textbf{b-d} for the $y$-valleys. The amplitude of the valley signal in-between the two donors is reduced compared to the sum of the two envelope as the $y$-valleys interfere destructively, as highlighted by the grey arrows.}
\label{figS1bis}
\end{figure}

\subsection{Comparison to FCI theory}

The theoretical STM image calculations protocol is described in the Methods section of the paper. The tunnelling matrix element between donor and tip states is defined as:

\begin{equation}
M=\frac{2}{3}  \frac{\partial^2 \Psi_D(r)}{\partial z^2}-\frac{1}{3}  \frac{\partial^2 \Psi_D(r)}{\partial y^2}-\frac{1}{3}  \frac{\partial^2 \Psi_D(r)}{\partial x^2}
\end{equation}

For two-particle STM images, two-electron wave functions are computed from a full configuration interaction approach~\cite{Tankasala2018}. The STM image represents a quasi-particle wave function resulting from $2e$ to $1e$ transition (see below). The resulting quasi-particle state is used to compute tunnelling matrix element described above. The exchange interaction obtained from FCI (see below) is the energy difference between the ground and the first triplet state. The FCI $2e$ ground state is defined as a sum of Slater determinants:

\begin{equation}
\ket{\Psi_s}=\sum_{\alpha \beta}^{}d_{\alpha \beta}\ket{\alpha \beta} \quad with \quad \ket{\alpha \beta}=\frac{1}{\sqrt{2}}(\ket{\alpha}_1\ket{\beta}_2-\ket{\beta}_1\ket{\alpha}_2)=c_{\alpha}^{\dagger}c_{\beta}^{\dagger}\ket{0}
\end{equation}

and $\mathrm{\ket{\alpha}}$ are $1e$ molecular orbital (including spin). The tunnelling current, governed by $\mathrm{\Gamma_{out} \ll \Gamma_{in}}$,  can be expressed as the sum of the $2e$ to $1e$ transitions following~\cite{Rontani2005}.

\begin{equation}
\Gamma_{out}(\vec{r})=\sum_i|M_{iS}(\vec{r})|^2=\sum_{i}|\bra{i}\Psi (\vec{r})\ket{\Psi_s}|^2
\end{equation}

with $\mathrm{\ket{i}=c_i^{\dagger}\ket{0}}$ and $\mathrm{\ket{\Psi(\vec{r}}=\sum_i c_i \phi_i(\vec{r})}$ the tunneling field operator. We obtain:

\begin{equation}
\begin{gathered}
M_{iS}(\vec{r})=\sum_{\alpha \beta}d_{\alpha \beta}\bra{0}c_i\sum_{j}c_j\phi_jc_{\alpha}^{\dagger}c_{\beta}^{\dagger}\ket{0}\\
M_{iS}(\vec{r})=\sum_{\alpha \beta}d_{\alpha \beta}(\phi_{\alpha}\delta_{i\beta}-\phi_{\beta}\delta_{i\alpha})
\end{gathered}
\end{equation}

The theoretical image is shown in Fig~\ref{figS1}e-f. We can observe that the valley pattern symmetries for each donor are well preserved. The 2D Fourier transform of this image, shown in Fig3a of the main text also shows the valley signal at $k_{x,y}\sim0.81*k_0$ and the same stripes as for the experimental data.\\

Further details on the FCI calculations for this specific pair are given in Fig~\ref{figS1bis}a. A charging energy of 34.3\,meV and an exchange energy (taken as the energy difference between the first singlet (ground) state and the first triplet state) of 1.5\,meV can be extracted from FCI calculations. Moreover, the FCI calculations gives the contribution of each molecular orbital in the $2e$ states. The FCI calculation singlet state yields a 67\% contribution from the $A_1$ bonding state (called $A_{1-B}$), 19\% contribution from the $A_1$ anti-bonding state (called $A_{1-AB}$), the rest coming from higher valley and orbital states, notably the $T_2$ bonding state which comes down in energy because of the tunnel coupling between the two donors~\cite{Rahman2011,Saraiva2015}. We can discuss the deviation to a pure Heitler-London state $2e$ molecular state which would be made of the single donor $A_1$ ground states. The net dominant contribution of the $A_1$ bonding state for pair \#1 comes for the finite tunnel coupling value: a pure HL state would result in a 50\% contribution of both $A_1$ bonding and anti-bonding states, as these two orbitals are degenerate for vanishing tunnel coupling. A similar study can be done for the triplet case, which shows a 80\% contribution of a combination of bonding and anti-bonding molecular orbitals (called $A_{1-B/AB}$) to form the triplet states. Finally, we show in Fig~\ref{figS1bis}d and g a line cut taken across the two ions for both the $x$ and the $y$-valley theoretical image in Fig~\ref{figS1bis}b and e, respectively, as well as of their respective 2D fit, Fig~\ref{figS1bis}c and f. Similarly to the experimental data, the $y$-valley signal in-between the two donors is weaker than the sum of the envelopes (grey dashed line in Fig~\ref{figS1bis}g) because of the destructive $y$-valley interference.

\subsection{STM image and Heitler-London regime}

Let's consider only the two $A_1$ ground states for each donor. The 2e ground state can be expressed as a combination of even and odd combination of $A_1$:

\begin{equation}
\begin{gathered}
\psi_S^{A1}(\vec{r})=\gamma_{ee}\phi_{ee}(\vec{r})+\gamma_{oo}\phi_{oo}(\vec{r}) \\
with \quad \phi_{ee}(\vec{r})=\frac{\psi_1^{A1}(\vec{r})+\psi_2^{A1}(\vec{r})}{c_{ee}\sqrt{2}} \quad and \quad \phi_{oo}(\vec{r})=\frac{\psi_1^{A1}(\vec{r})-\psi_2^{A1}(\vec{r})}{c_{oo}\sqrt{2}}
\end{gathered}
\end{equation}

The associated expression for the tunneling current follows:
\begin{equation}
\Gamma_{out}(\vec{r})=|\gamma_{ee}|^2|\phi_{ee}(\vec{r})|^2+|\gamma_{oo}|^2|\phi_{oo}(\vec{r})|^2
\end{equation}

In the Fermi-Hubbard framework the Heitler-London limit $\mathrm{U/t \rightarrow \infty}$ leads to $\mathrm{\gamma_{ee}=\gamma_{oo}=1/\sqrt{2}}$~\cite{Salfi2016}. This results in the following STM current:

\begin{equation}
\begin{gathered}
\Gamma_{out}^{HL}(\vec{r}) \propto |\phi_{ee}(\vec{r})|^2+|\phi_{oo}(\vec{r})|^2 \propto |\psi^{A1}_1(\vec{r})+\psi^{A1}_2(\vec{r})|^2 +|\psi^{A1}_1(\vec{r})-\psi^{A1}_2(\vec{r})|^2 \\
\Gamma_{out}^{HL}(\vec{r}) \propto |\psi^{A1}_1(\vec{r})|^2+|\psi^{A1}_2(\vec{r})|^2
\end{gathered}
\end{equation}

which is a $\mathrm{|D_0}(\vec{r}-\vec{r_1})|^2+\mathrm{|D_0}(\vec{r}-\vec{r_2})|^2$ STM image, i.e. the sum of the images of two single donors. This limit makes an evident link between the $2e$ STM image and the $1e$ probability density for each donor, which hence contain both the geometric valley interference terms between the two donors. As it can also be deduced between the Heitler-London relationship $J=4t^2/U$ between the tunnel coupling $t$ (a $1e$ quantity) and the exchange coupling $J$ (a $2e$ quantity), valley interference impact both $1e$ and $2e$ processes.

\subsection{2nd pair valley analysis}

\begin{figure}[htp]
\includegraphics{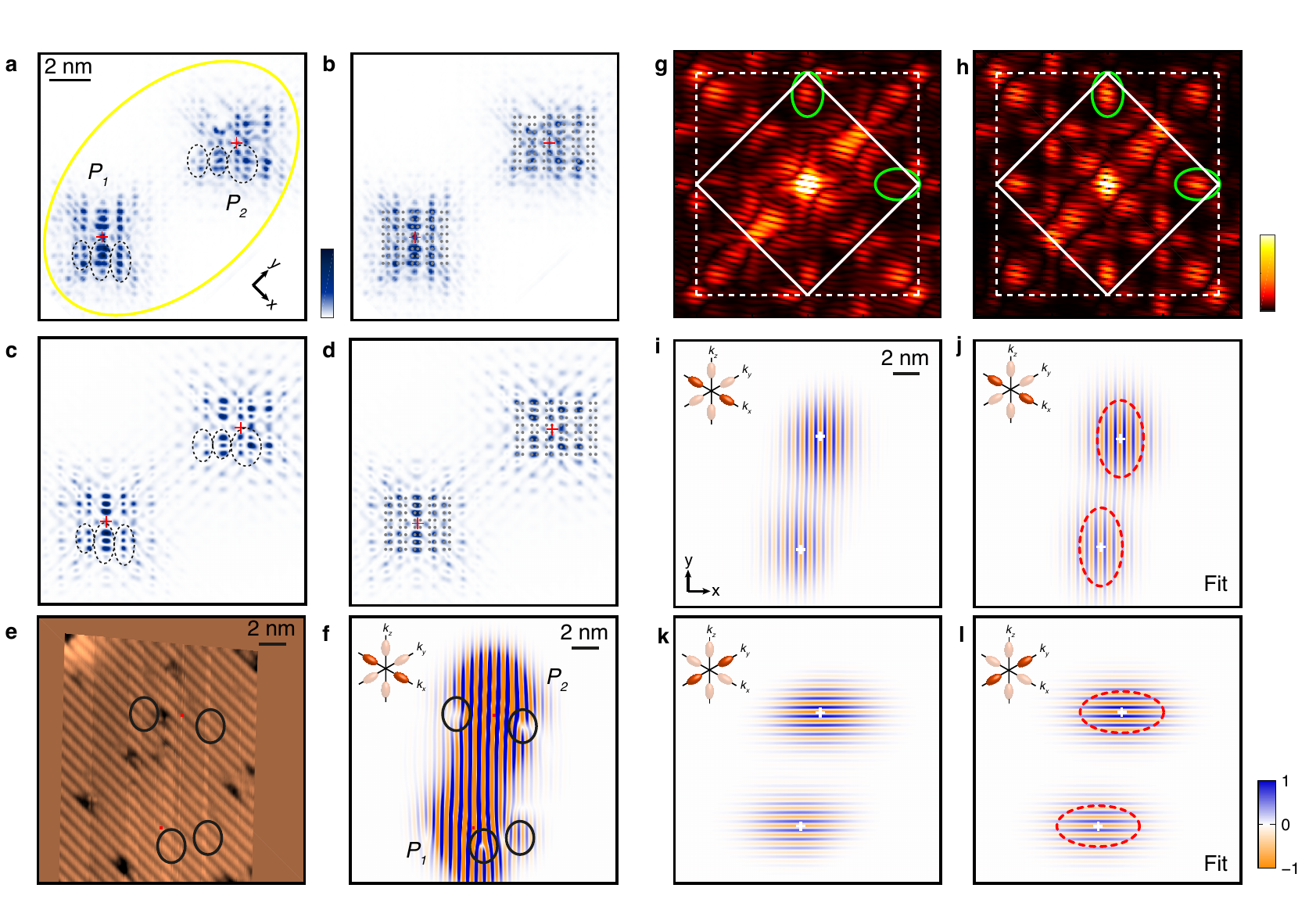}
\caption{ \textbf{Valley analysis for pair \#2. a,} Experimental STM image. The yellow ellipse represents the contour filter used before the Fourier transform. \textbf{b,} Experimental image with the surface lattice superimposed. Donor $P_1$ sits in the middle of a dimer row, in-between two pairs of atoms forming a dimer. Donor $P_2$ sits underneath a row of silicon atoms, in-between two dimer pairs. \textbf{c,} Corresponding theoretical STM image calculated from a Heitler-London $2e$ states, i.e. a $\rm{|D_0|^2+|D_0|^2}$ image. The black ellipses in \textbf{a} and \textbf{c} point to instances of matching features between the two images which were used to pinpoint the donors location. \textbf{d,} Same as \textbf{c} with the surface silicon lattice superimposed. \textbf{e,} Experimental topography taken at $U{=}-1.6$\,V. There are evident tip jumps circled in black in the measurement along the $x$-axis (slow scan). \textbf{f,} 2D $x$-valley image, same as main text only with a saturated color scale, in order to focus on the forks and distortions of the valley oscillations. The black circles from \textbf{e} placed at the same locations evidence for the tip jumps to be origin of the distortions in the valley oscillations. \textbf{g,} Experimental Fourier transforms based on \textbf{a}. \textbf{h,} Theoretical Fourier transform based on \textbf{c}. \textbf{i,} Resulting $x$-valley image from Fourier filtering around the conduction band minima with $f_r{=}0.1(2\pi/a_0)$ and $f_t{=}0.7f_r$. \textbf{j,} Fit of the $x$-valley image using eq.~\ref{fiteqx}. \textbf{k,l} Same as  \textbf{i,j} for the $y$-valleys. The fit parameters shown in the tables of the main text  for pair \#2 originate from these images.}
\label{figS2HL}
\end{figure}

Here we present complimentary data for pair \#2, where the donors are found to be distant by $\rm{13a_0\sqrt{2}/2}$ along [110] and $\rm{9.25a_0\sqrt{2}}$ along $\rm{[1\bar{1}0]}$. Donor $P_1$ is found at $z{=}6.5a_0$ and $P_2$ at $z{=}6.25a_0$. We show the surface lattice superimposed on the STM image in Fig.~\ref{figS2HL}b. The yellow contour shown in Fig.~\ref{figS2HL}a shows the filter applied on the data before the Fourier transform. For this pair, the theoretical image was computed using the Heitler-London limit for the 2e ground state as reference, which is relevant for the inter-donor distance $d{>}7$\,nm measured in this case. The resulting theoretical image STM image is shown in Fig.~\ref{figS2HL}c and d, without and with the lattice superimposed, respectively. We note the excellent agreement between the features of the images along and across the dimers between the experimental and theoretical images, which are used to pinpoint the donors location~\cite{Usman2016}, as it was done for pair \#1. The Fourier transforms of the images are shown in Fig.~\ref{figS2HL}g and h, for the experimental and the theoretical image, respectively. Again, we note a good agreement on the features present in the first Brillouin zone, with only a weaker signal for the $x$-valleys in the case of the experiment, which we believe is due to the jumps in the tunnel current while the STM image was being measured with a slow scan along this $x$-axis. These jumps are best seen in the topography pass, as shown in Fig.~\ref{figS2HL}e. The black circles on this topographic image run over some of these tip jump, and they match the forks, or distortions, seen in the $x$-valley image of the experimental data shown in the main text, and reproduced and amplified with a saturated colorscale here in Fig.~\ref{figS2HL}f. To complement the data shown in the main text regarding pair \#2, we show here in Fig.~\ref{figS2HL}i-l the $x$ and $y$-valley images as well as their fit for the theoretical STM image, which correspond to the phase differences shown in Fig.5 of the main text and to the envelope radius anisotropy shown in the table below.\\

\subsection{Robustness of the valley phase difference to the dimension of the Fourier filter.}

\begin{figure}[htp]
\hspace{-0.05cm}
\includegraphics{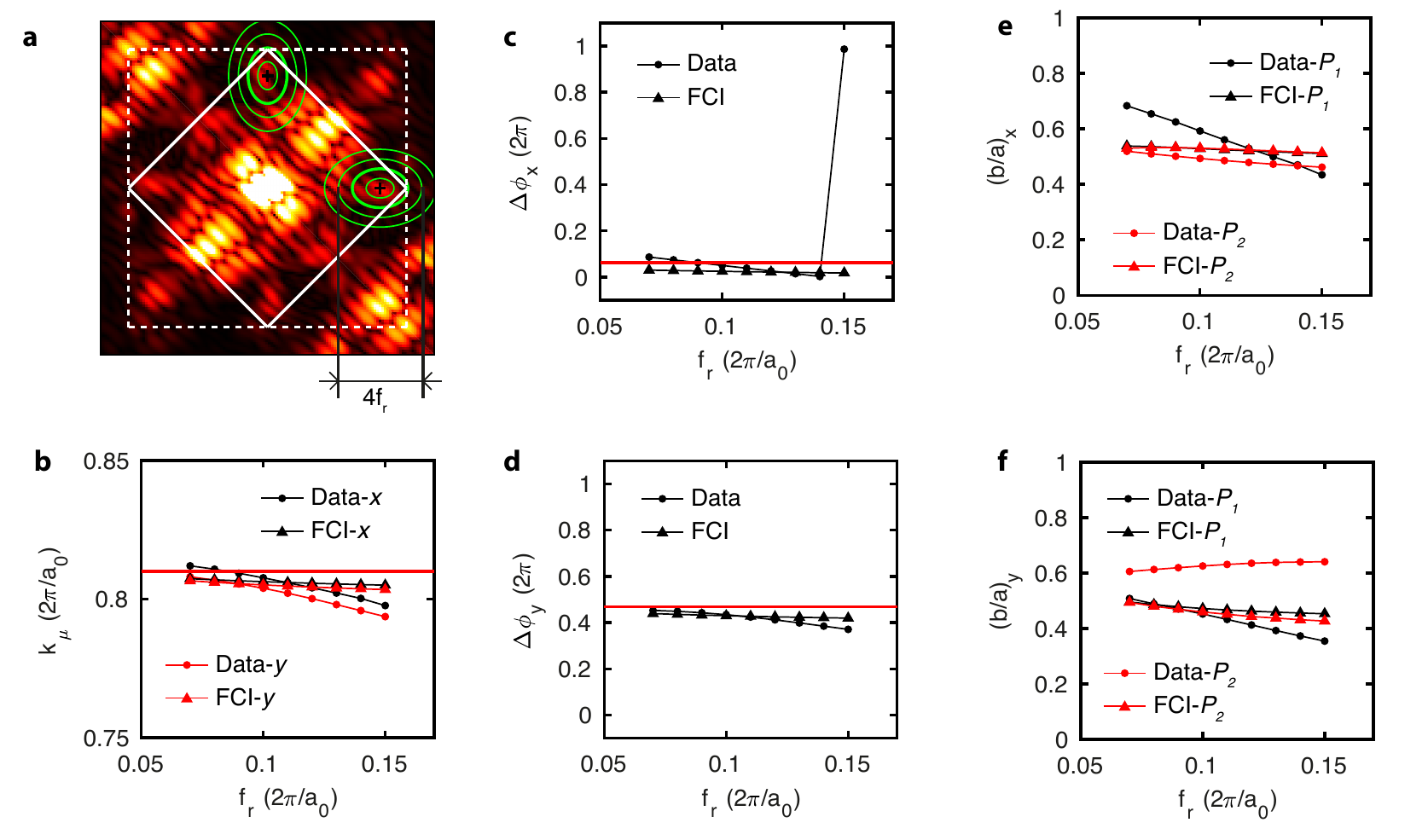}
\caption{ \textbf{Analysis of the fit parameters vs filter dimension for pair \#1. a,} FFT of the experimental STM image. The fitting procedure was performed for different Fourier filter dimensions $f_r$ values in \ref{maskx} and \ref{masky}, with $f_t{=}0.7f_r$. $f_r{=}\{0.05,0.10,0.15,0.20\}(2\pi/a_0)$ from the inner to the outer green ellipsoids around the valley signal at $k_{x,y}=0.81k_0$, respectively. \textbf{b,} Valley momentum obtained vs $f_r$ for both the experiment (dots) and FCI calculations (triangles), and both $x$ (black) and $y$-valleys. The red line corresponds to $k_{\mu}{=}0.81k_0$. The values change by less than 2\% over the range of $f_r$ which was studied. \textbf{c,} $\Delta\phi_x$ vs $f_r$ for the experiment and FCI. The red line at $0.06(2\pi)$ corresponds to the predicted geometric interference condition from the donors lattice position and  $k_{\mu}{=}0.81k_0$. Both  experimental and theroretical values remain very close to the geometrical interference condition. The experimental point for $f_r{=}0.15(2\pi/a_0)$ is simply shifted by a modulo $[2\pi]$. \textbf{d,} Same for $\Delta\phi_y$. \textbf{d,} Anisotropy $b/a$ obtained for the $x$-valleys, for both experiment (dots) and FCI (triangles), for both $P_1$, i.e. $b_1/a_1$ in eq.~\ref{fiteqx} (black) and $P_2$, i.e. $b_2/a_2$ in eq.~\ref{fiteqy} (red). The experimental anisotropy for $P_1$ shows a larger dependence with $f_r$, while the others are close to the effective mass value of 0.52. \textbf{e,} Same for for the $y$-valleys.}
\label{figS1fit}
\end{figure}

Choosing an appropriate filter in Fourier space is crucial for our analysis, and a trade-off must be considered. On one hand, the filter should not be too small as it would restrict the range of frequencies which could be obtained and enforce a single value. Moreover, using a filter whose $k$-space extent is smaller than the feature of interest would result in an artificially enlarged feature in real-space, and the result envelop radii and anisotropy could not be trusted. On the other hand, the filter cannot be too large as it would eventually include other Fourier components than that of interest, which would alter the fitting procedure in real space and the relevance of the results. We show in Fig.~\ref{figS1fit}a the FFT for pair \#1 and a range of different ellipsoids around the $x$ and $y$-valley components, with respectively $f_r=\{0.05,0.1,0.15,0.2\}(2\pi/a_0)$ from the inner to the outer ellipsoid, and $f_t/f_r{=}0.7$. Clearly the ellipsoid with $f_r{=}0.05(2\pi/a_0)$ does not englobe the whole valley signal, while $f_r{=}0.15(2\pi/a_0)$ leaks out to other components of the Fourier image. Therefore we estimate that a range from $f_r{=}0.07(2\pi/a_0)$ to $f_r{=}0.15(2\pi/a_0)$ is appropriate to analyse this valley component. The figures presented in the main text correspond to images and fitting parameters all obtained with $f_r{=}0.10(2\pi/a_0)$ and $f_t{=}0.7f_r$.\\

We show in Fig.~\ref{figS1fit}b the evolution of $k_{\mu}$ with $f_r$, for both experimental and FCI data, and both $x$ and $y$-valleys. A small dependence can be observed for the experimental values, however it is remarkable that all the values are found to vary by less than 2\% over the considered range of $f_r$, which is extremely restrained considering the range of frequencies which has become available for $f_r{=}0.15(2\pi/a_0)$. The resulting dependence for $\Delta\phi_x$ and $\Delta\phi_y$ are shown in Fig.~\ref{figS1fit}c-d, respectively. The phase differences remain very close to the predicted geometric interference, with also very limited spread which can be related to the dependence in $k_{\mu}$. The ratio in the envelope radii, shown in Fig.~\ref{figS1fit}e-f, fir the $x$ and $y$-valleys respectively, consistently remains below 0.65, which evidences a clear anisotropy. We note that the $x$-valleys of $P_1$ shows a stronger dependence of the ratio with the filter dimension than the other values. We have performed the same procedure for pair \#2, with the results shown in Fig.~\ref{figS2Fit}. We obtain very similar robustness of the phase differences anisotropy as for pair \#1. We note that an enhanced dependence of $k_x$ with $f_r$, although still limited to 2\%, which we attribute to the instability of the tunneling tip as explained above.\\

\begin{figure}[htp]
\includegraphics{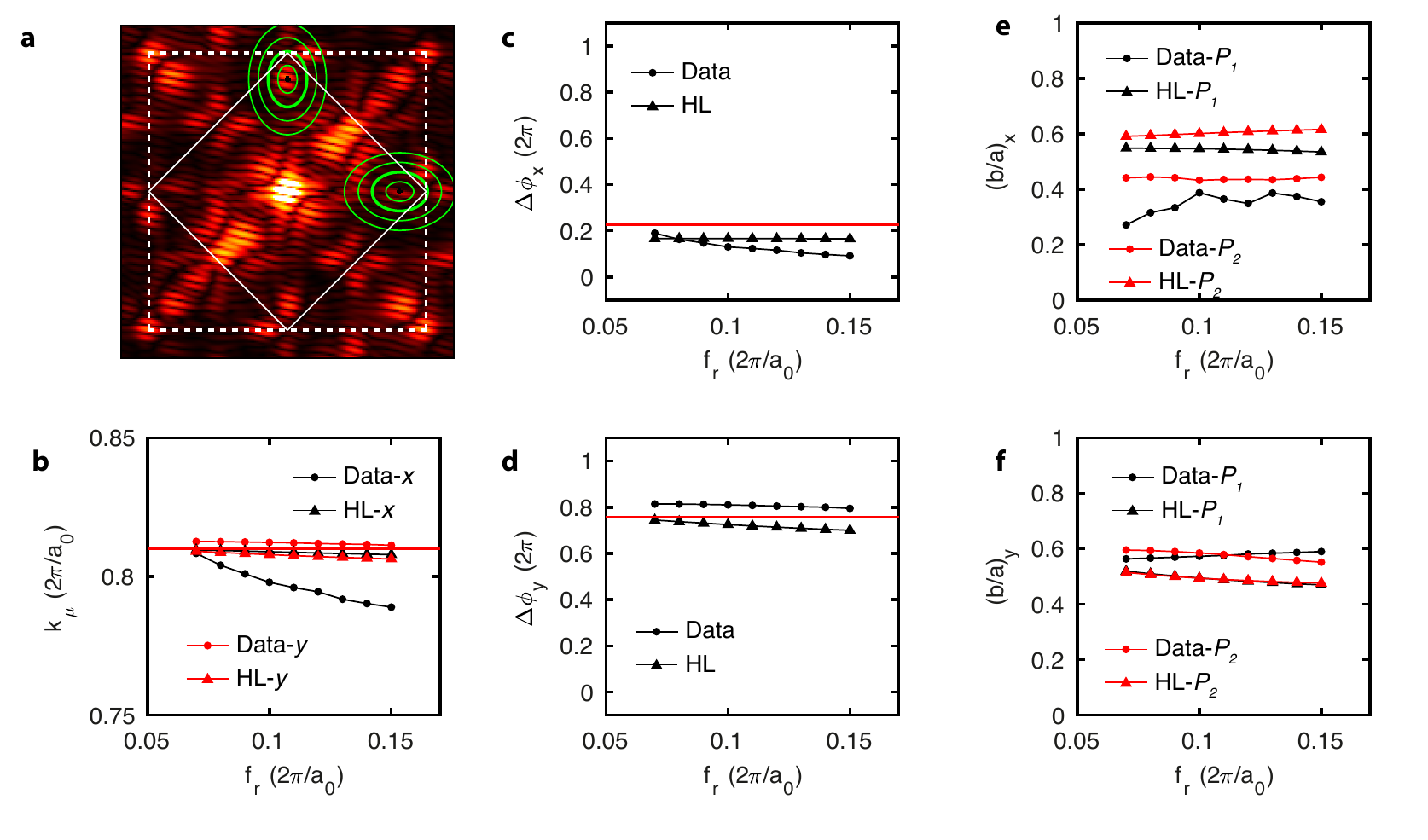}
\caption{ \textbf{Analysis of the fit parameters vs filter dimension for pair \#2. a,} FFT of the experimental STM image. The fitting procedure was performed for different Fourier filter dimensions $f_r$ values in \ref{maskx} and \ref{masky}, with $f_t{=}0.7f_r$. $f_r{=}\{0.05,0.10,0.15,0.20\}(2\pi/a_0)$ from the inner to the outer green ellipsoids around the valley signal at $k_{x,y}=0.81k_0$, respectively. \textbf{b,} Valley momentum obtained vs $f_r$ for both the experiment (dots) and FCI calculations (triangles), and both $x$ (black) and $y$-valleys. The red line corresponds to $k_{\mu}{=}0.81k_0$. The 2\% deviation of experimental value for $k_x$ from $0.81k_0$ is attributed to the tip jumps associated with the measurement. \textbf{c,} $\Delta\phi_x$ vs $f_r$ for the experiment and FCI. The red line at $0.23(2\pi)$ corresponds to the predicted geometric interference condition from the donors lattice position and $k_{\mu}{=}0.81k_0$. \textbf{d,} Same for $\Delta\phi_y$. \textbf{d,} Anisotropy $b/a$ obtained for the $x$-valleys, for both experiment (dots) and FCI (triangles), for both $P_1$, i.e. $b_1/a_1$ in eq.~\ref{fiteqx} (black) and $P_2$, i.e. $b_2/a_2$ in eq.~\ref{fiteqy} (red). The values, ranging from 0.4 to 0.6 apart from the experimental value for $P_1$ due to the tip jumps, are close to the effective mass value of 0.52. \textbf{e,} Same for for the $y$-valleys. Again the values are on average close to 0.52.}
\label{figS2Fit}
\end{figure}

\clearpage

\subsection{Envelope anisotropy}

We show in Fig.~\ref{figS1envelope} the anisotropy ratios $b/a$, as well as standard deviations, obtained from fitting the 2D filtered images according to equations~\ref{fiteqx} and ~\ref{fiteqy}. Each donor of each pair gives a value for the $x$ and the $y$-valleys, for both experimental and theoretical images, which results in a total of 16 values. The values show a spread from 0.387 to 0.625, which clearly establishes the existence of an envelope anisotropy. The values average to 0.52 in good agreement with single donors measurements~\cite{Saraiva2016}.\\

Finally, we discuss the dimensions of the donor's envelope obtained from the fits with respect to the dimension of the Fourier filter. As shown in equations~\ref{maskx} and ~\ref{masky}, the filters have a variance equal to $f_r(2\pi/a_0)$ (respectively $f_t(2\pi/a_0)$ ) in the longitudinal (transverse) direction. Back to real-space, this is obviously equivalent to a characteristic dimension $a_0/ (2\pi f_r)$ in the longitudinal direction, which must be compared to the longitudinal envelope radii $b$,  and to $a_0/ (2\pi f_t)$ in the transverse direction to be compared with $a$. We consider here the values which correspond to $f_r{=}0.10(2\pi/a_0)$, as they are the ones presented in the main text. Over 16 values (2 valleys, 2 pairs, 2 donors/pair, experimental and theoretical images), the small envelope radii $b$ range from 1.14 to 1.85\,nm, with an average of 1.52\,nm, which is much larger than the filter longitudinal dimension equal to 0.84\,nm. Likewise, the large envelope radii $b$ range from 2.34 to 3.48\,nm, with an average of 2.96\,nm, which is also much larger than the filter's transverse dimension equal to 1.23\,nm. This demonstrates the relevance of our fitting procedure to accurately extract not only the phase differences but also the envelope parts of the valley images discussed throughout the manuscript. Importantly, the overlap between the envelopes can be confidently discussed, to further prove the robustness of the valley interference in these regions, as well as the anisotropy of the envelope longitudinal and transverse dimensions. These are the two core ingredients of the effective mass model used to discuss their impact on the exchange interaction.\\ 

\begin{figure}[htp]
\hspace{-0.05cm}
\includegraphics{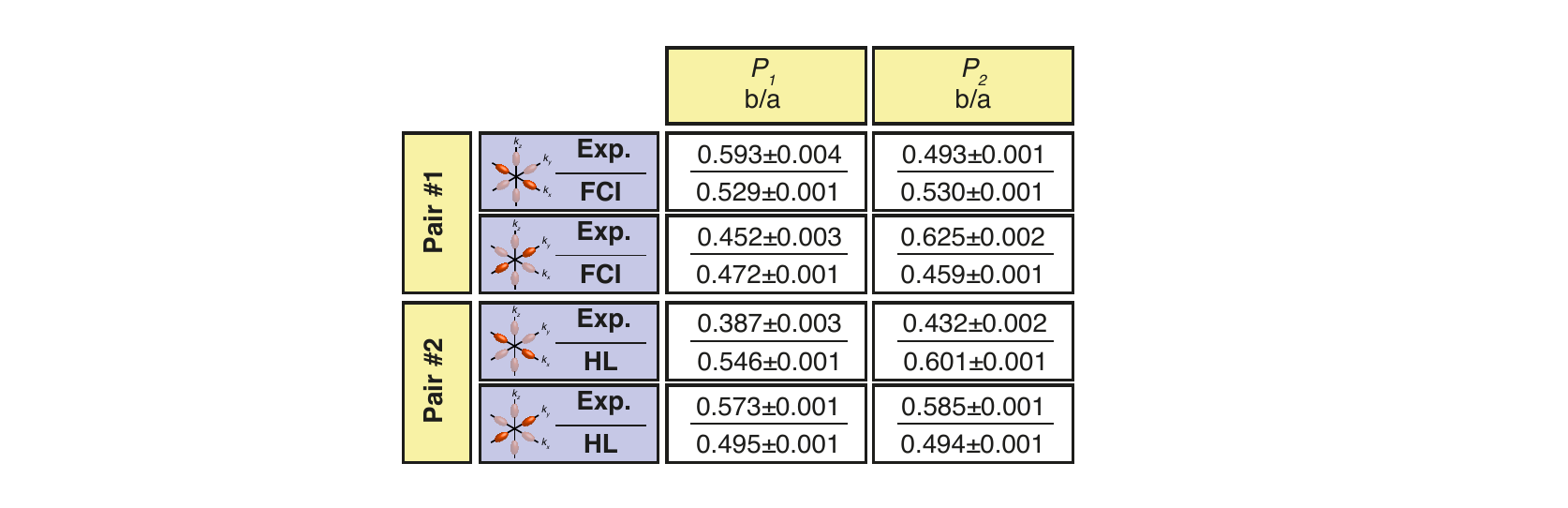}
\caption{\textbf{Envelope anisotropy fitting results.} Table summarising the anisotropy ratios $b/a$ for each donor, obtained from the fit of the valley images for the two pairs, for both experimental and theoretical images. The values average to 0.52 in agreement with single donors measurements and calculations.}
\label{figS1envelope}
\end{figure}

\clearpage

\section{Supplementary Note 2 - Spectroscopy of exchange-coupled donors}

We give in Fig.~\ref{figS32} the spectroscopic data measured for pair \#1 of the main text. Fig.~\ref{figS32}b shows the differential conductance vs $U$ along a cut though the two donors. The different charge state transitions can be identified as the main resonance peaks occurring around -0.5V, -0.8V and -1.15V (black arrows in Fig.~\ref{figS32}b-c), for the 0e/1e, 1e/2e and 2e/3e, respectively, with the neutral 1e/2e state transition occurring close to the flat-band condition (zero electric field) at -0.8V because of tip-induced band bending~\cite{Salfi2014,Voisin2015,Salfi2018}. We also note that the 0e/1e transition is mainly located on the bottom donor from Fig.~\ref{figS32}a. This might be due to a combination of stray electric field from the environment as well as electric field created by the tip at this repulsive bias value.\\

As mentioned, we do not expect the local tip-induced stray field to distort the 2e image because this image is taken at a bias close to the flat-band condition, i.e. zero electric field. More importantly, we expect the valley phase to be robust against electric fields from the tip, because the valley frequency is not sensitive to electric field and, even in the presence of a moderate Stark shift, the phase is always pinned to the donor's ion position. 

\begin{figure}[htp]
\hspace{-0.05cm}
\includegraphics{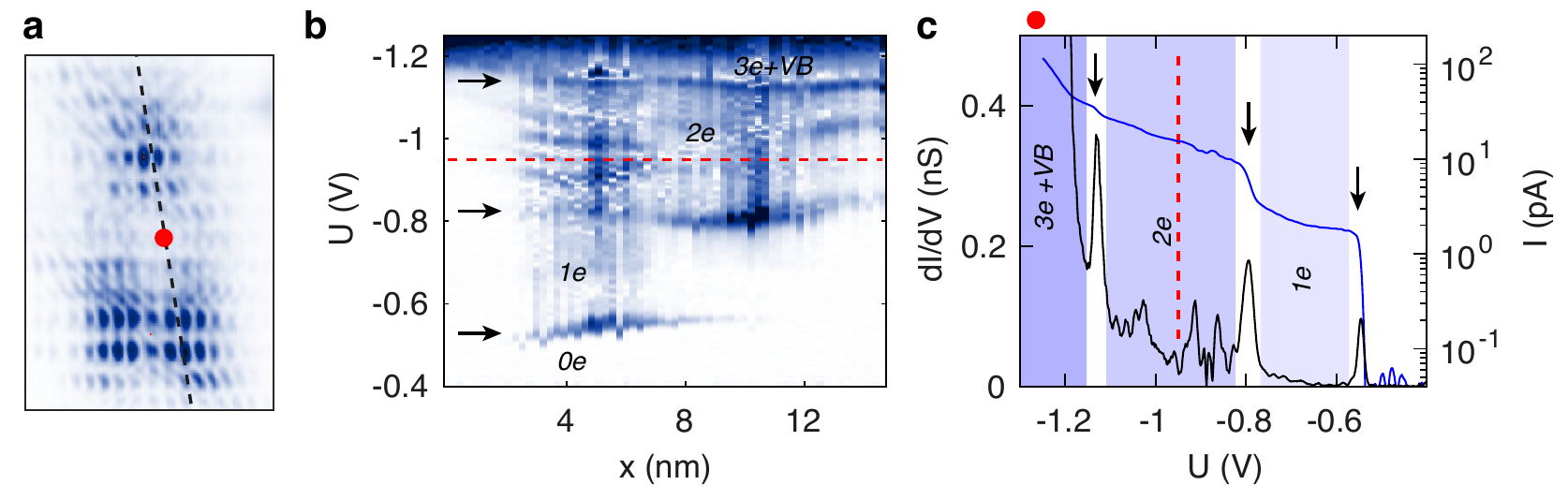}
\caption{ \textbf{Spectroscopy data for the pair presented in the main text. a,} STM image of the pair discussed in the main text. \textbf{b,} Map of the differential conductance plotted vs bias voltage taken along the black dotted line shown in a. The different charge states (0e, 1e, 2e and 3e) can be identified from the main differential conductance peaks around, respectively, -0.5V, -0.8V and -1.15V, indicated by the black arrows. The red doted line indicates the bias voltage at which the STM image was taken, above the 1e/2e transition. \textbf{c,} Differential conductance plotted vs bias voltage taken at the red spot shown in a. The different charge state transitions can clearly be identified (black arrows).}
\label{figS32}
\end{figure}

\newpage

\section*{Supplementary Note 3 - Valley interference, STM placed donors and exchange analysis}

\subsection{Heitler-London exchange for two donors in silicon.}

In this section we derive the expression of the exchange interaction energy between two donors in the Heitler-London (HL) regime~\cite{Koiller2001}. For convenience we start from a single k-point, A1-like donor ground state:

\begin{equation}
\begin{gathered}
\psi(\vec{r})=\sum_{\mu =1}^6 \alpha_{\mu}F_{\mu}(\vec{r})\phi_{\mu}(\vec{r}) \\
with \quad F_{\pm z}(\vec{r})=\frac{1}{\sqrt{\pi a^2b}}e^{-\sqrt{\frac{(x^2+y^2)}{a^2}+\frac{z^2}{b^2}}} \quad \textrm{anisotropic \; envelope}\\
and \quad \phi_{\mu}(\vec{r})=u_{\mu}(\vec{r})e^{i\vec{k_{\mu}}\cdot \vec{r}} \quad \textrm{Bloch \; functions}
\end{gathered}
\label{EMenv}
\end{equation}

The $\mathrm{\alpha_{\mu}}$ represent the valley population distribution among the 6 valleys $\pm x$, $\pm y$ and $\pm z$ and are assumed to be real numbers. $a$ and $b$ represent the transverse and longitudinal donor envelope radii, respectively. The average ratio $b/a{=}0.52$ obtained in this work is in good agreement with reported values for single donors~\cite{Saraiva2016}, the value $a$ is discussed below in the context of atomistic calculations of the exchange interaction.\\

The HL-exchange interaction energy is defined as:
\begin{equation}
\begin{gathered}
J(\vec{R})=\iint \vec{dr_1}\vec{dr_2}\psi^*(\vec{r_1})\psi^*(\vec{r_2}-\vec{R})\frac{e^2}{\epsilon |\vec{r_1}-\vec{r_2}|}\psi(\vec{r_1}-\vec{R})\psi(\vec{r_2})\\
=\iint \vec{dr_1}\vec{dr_2} \Big(\sum_{\mu}\alpha_{\mu}F_{\mu}^*(\vec{r_1})\phi_{\mu}^*(\vec{r_1})\Big) \Big(\sum_{\nu}\alpha_{\nu}F_{\nu}^*(\vec{r_2}-\vec{R})\phi_{\nu}^*(\vec{r_2}-\vec{R}) \Big)\frac{e^2}{\epsilon |\vec{r_1}-\vec{r_2}|}\times \\
\Big(\sum_{\mu '}\alpha_{\mu '}F_{\mu '}(\vec{r_1}-\vec{R})\phi_{\mu '}(\vec{r_1}-\vec{R})\Big) \Big(\sum_{\nu '}\alpha_{\nu'}F_{\nu'}^*(\vec{r_2})\phi_{\nu'}^*(\vec{r_2}) \Big)\\
=\iint \vec{dr_1}\vec{dr_2} \sum_{\mu \nu \mu ' \nu '} \alpha_{\mu}\alpha_{\nu}\alpha_{\mu '}\alpha_{\nu '} F_{\mu}(\vec{r_1})^*F_{\nu}(\vec{r_2}-\vec{R})^*\frac{e^2}{\epsilon |\vec{r_1}-\vec{r_2}|}F_{\mu '}(\vec{r_1}-\vec{R})F_{\nu '}(\vec{r_2}) \times \\
 \phi_{\mu}(\vec{r_1})^* \phi_{\nu}(\vec{r_2}-\vec{R})^* \phi_{\mu '}(\vec{r_1}-\vec{R}) \phi_{\nu '}(\vec{r_2})
\end{gathered}
\end{equation}

We define the envelope part $\tilde{j}_{\mu \nu \mu ' \nu '}=F_{\mu}^*(\vec{r_1})F_{\nu}^*(\vec{r_2}-\vec{R})\frac{e^2}{\epsilon |\vec{r_1}-\vec{r_2}|}F_{\mu '}(\vec{r_1}-\vec{R})F_{\nu '}(\vec{r_2})$. We develop the Bloch function parts using:

\begin{equation}
u_{\mu}(\vec{r})=\sum_{\vec{K}}c_K^{\mu}e^{i\vec{K}\cdot \vec{r}} \quad with \quad \sum_{\vec{K}}|c_K^{\mu}|^2=1
\end{equation}

We then obtain:
\begin{equation}
\begin{gathered}
J(\vec{R})=\sum_{\mu \nu \mu ' \nu '} \alpha_{\mu}\alpha_{\nu}\alpha_{\mu '}\alpha_{\nu '} \iint \vec{dr_1}\vec{dr_2}
\tilde{j}_{\mu \nu \mu ' \nu '}e^{i(\vec{k_{\mu '}}-\vec{k_{\mu}})\cdot \vec{r_1}}e^{i(\vec{k_{\nu '}}-\vec{k_{\nu}})\cdot \vec{r_2}} e^{i(\vec{k_{\nu}}-\vec{k_{\mu '}})\cdot \vec{R}} \times \\
\sum_{\vec{K_1}..\vec{K_4}}c_{\vec{K_1}}^{\mu *}c_{\vec{K_2}}^{\nu *}c_{\vec{K_3}}^{\mu '}c_{\vec{K_4}}^{\nu '}e^{-i\vec{K_1}\cdot \vec{r_1}}e^{-i\vec{K_2}\cdot (\vec{r_2}-\vec{R})}e^{i\vec{K_3}\cdot (\vec{r_1}-\vec{R})}e^{i\vec{K_4}\cdot \vec{r_2}}
\end{gathered}
\end{equation}

A few assumptions are then made to obtain the final expression. First we neglect the fast oscillating terms in (S9) of the form $e^{i(\vec{k_{\mu '}}-\vec{k_{\mu}})\cdot \vec{r_1}}$ with $\mathrm{\mu ' \neq \mu}$ and likewise $e^{i(\vec{k_{\nu '}}-\vec{k_{\nu}})\cdot \vec{r_2}}$ with $\mathrm{\nu ' \neq \nu}$, which are called the inter-valley exchange terms as the electrons change valley and site during exchange. These fast oscillating terms would integrate to zero for inter-donor distances larger than the envelope radius~\cite{Koiller2001,Salfi2016} which we consider in this work. Therefore, we neglect them in the following and only consider intra-valley processes. Hence $\mathrm{\mu = \mu ' \rightarrow \mu}$ for electron 1 and $\mathrm{\nu = \nu ' \rightarrow \nu}$ for electron 2. Assuming $\mathrm{\alpha_{\mu} =\alpha_{\mu '}}$ and $\mathrm{\alpha_{\nu} =\alpha_{\nu '}}$ the expression for the exchange becomes:

\begin{equation}
\begin{gathered}
J(\vec{R})=\sum_{\mu \nu} \alpha_{\mu}^2\alpha_{\nu}^2\iint \vec{dr_1}\vec{dr_2}\tilde{j}_{\mu \nu} e^{i(\vec{k_{\nu}}-\vec{k_{\mu}})\cdot \vec{R}}\sum_{\vec{K_1}..\vec{K_4}}c_{\vec{K_1}}^{\mu *}c_{\vec{K_2}}^{\nu *}c_{\vec{K_3}}^{\mu}c_{\vec{K_4}}^{\nu}e^{i(\vec{K_3}-\vec{K_1})\cdot \vec{r_1}}e^{i(\vec{K_4}-\vec{K_2})\cdot \vec{r_2}}e^{i(\vec{K_2}-\vec{K_3})\cdot \vec{R}}\\
=\sum_{\mu \nu} \alpha_{\mu}^2\alpha_{\nu}^2\iint \vec{dr_1}\vec{dr_2}\tilde{j}_{\mu \nu} e^{i(\vec{k_{\nu}}-\vec{k_{\mu}})\cdot \vec{R}}\sum_{\vec{K_1} \vec{K_2}} |c_{\vec{K_1}}^{\mu}|^2|c_{\vec{K_2}}^{\nu}|^2e^{i(\vec{K_1}-\vec{K_2})\cdot \vec{R}}
\end{gathered}
\end{equation}

$\vec{K_1}-\vec{K_2}$ is a reciprocal lattice vector and $\vec{R}$ is a lattice vector for substitutional donors, thus $e^{i(\vec{K_1}-\vec{K_2})\cdot \vec{R}}=1$. Using the normalisation condition for the Bloch states we finally obtain:

\begin{equation}
\begin{gathered}
J(\vec{R})=\sum_{\mu\nu} J_{\mu \nu}= \sum_{\mu \nu} \alpha_{\mu}^2\alpha_{\nu}^2 \; j_{\mu \nu}(\vec{R})\; \cos((\vec{k_{\mu}}-\vec{k_{\nu}})\cdot \vec{R})\\
with \quad j_{\mu \nu}(\vec{R})=\iint \vec{dr_1}\vec{dr_2}\tilde{j}_{\mu \nu}(\vec{R})=\iint \vec{dr_1}\vec{dr_2}F_{\mu}^*(\vec{r_1})F_{\nu}^*(\vec{r_2}-\vec{R})\frac{e^2}{\epsilon |\vec{r_1}-\vec{r_2}|}F_{\mu}(\vec{r_1}-\vec{R})F_{\nu}(\vec{r_2})
\label{jPEM}
\end{gathered}
\end{equation}

similar to ref~\cite{Koiller2001,Pica2014}.

\subsection{Phenomenological effective mass model}

This section describes how we constructed the phenomenological effective-mass model (P-EM). We first start from the general exchange equation (S11). For the bulk donor devices fabricated by STM lithography we use constant $\rm{\alpha_{\mu}=1/\sqrt{6}}$, hence taking the valley population weights out of the equation, as we will only be interested in exchange variations later on. It reduces the exchange expression to only a sum of envelope terms modulated by valley interference:

\begin{equation}
\begin{gathered}
J^{PEM}(\vec{R})= \sum_{\mu,\nu=1}^6 \; j_{\mu \nu}(\vec{R})\; \cos((\vec{k_{\mu}}-\vec{k_{\nu}})\cdot \vec{R})
\end{gathered}
\end{equation}

The terms $(\vec{k_{\mu}}-\vec{k_{\nu}})\cdot \vec{R}$ can be rewritten as $\Delta\phi_{\mu\nu}$ as used in the main text with $\Delta\phi_{\mu\nu}=\textrm{sign}(\mu)\Delta\phi_{\mu}-\textrm{sign}(\nu)\Delta\phi_{\nu}$ which makes the link with the valley phase differences experimentally measured. The envelope terms can be simplified assuming $F_{\mu}(\vec{r_1})F_{\nu}(\vec{r_2}) \sim F_{\mu}(\vec{r_1})F_{\nu}(\vec{r_2})\delta(\vec{r_1}-\vec{r_2})$ because of the exponential nature of the orbitals. This leads to:

\begin{equation}
\begin{gathered}
j_{\mu \nu}(\vec{R})=\iint \vec{dr_1}\vec{dr_2}F_{\mu}^*(\vec{r_1})F_{\nu}^*(\vec{r_2}-\vec{R})\frac{e^2}{\epsilon |\vec{r_1}-\vec{r_2}|}F_{\mu}(\vec{r_1}-\vec{R})F_{\nu}(\vec{r_2})\\
\sim \int \vec{dr_2}F_{\mu}^*(\vec{r_2}+\vec{R})F_{\nu}^*(\vec{r_2}-\vec{R})\frac{e^2}{\epsilon |\vec{R}|}F_{\mu}(\vec{r_2})F_{\nu}(\vec{r_2})\\
\sim \int \vec{dr_2}F_{\mu}^*(\vec{r_2}+\vec{R})F_{\nu}^*(\vec{R}-\vec{r_2})\frac{e^2}{\epsilon |\vec{R}|}F_{\mu}(\vec{r_2})F_{\nu}(\vec{r_2})\\
\sim \frac{1}{|\vec{R}|}F_{\mu}(\vec{R})F_{\nu}(\vec{R})F_{\mu}(\vec{0})F_{\nu}(\vec{0})=j_{\mu \nu}^{PEM}(\vec{R})\\
\label{envexpress}
\end{gathered}
\end{equation}

The products $F_{\mu}(\vec{R})F_{\mu}(\vec{0})$ (respectively $F_{\nu}(\vec{R})F_{\nu}(\vec{0})$) reflect the 3D exchange integrals for the electron exchanged in valley $\mu$ (respectively $\nu$) between the two donors, and still contain the anisotropy and exponential nature of the envelope integrals. For instance, as discussed in the main text, considering two donors along [100] and $\vec{R}=na_0\vec{x}$:

\begin{equation}
\begin{gathered}
F_{x}(na_0\vec{x})F_{x}(\vec{0})\ll F_{z}(na_0\vec{x})F_{z}(\vec{0}) \quad \Rightarrow \quad j_{xz}^{PEM}(na_0\vec{x}) \ll j_{zz}^{PEM}(na_0\vec{x})
\end{gathered}
\end{equation}

\subsection{An illustrative case - [110]}

The P-EM model developed above shows a complex 3D interplay between envelope anisotropic weights modulated by valley phase differences $\Delta\phi_{\mu}$. In order to illustrate their specific role, we can temporarily ignore any dopant misplacement and focus on the [110] direction only. Along this direction, we easily find from a symmetry argument  $\Delta\phi_{x}=\Delta\phi_{y}=\Delta\phi_{[110]}$ and $\Delta\phi_{z}=0$. The phase difference $\Delta\phi_{[110]}$ is plotted in Fig.~\ref{figS43}a vs the inter-donor distance. A clear qualitative correlation arises between $\Delta\phi_{[110]}$ (wrapped between 0 and $\pi$) and the exchange energy computed from FCI, plotted in Fig.~\ref{figS43}b: destructively interfering positions, i.e. $\Delta\phi_{[110]}$ close to $\pi$, match local minima in the exchange, with local variations of one to two orders of magnitude. This remains valid on a large range of distance, from a few nanometers where the exchange energy compares with the valley-orbit splitting and results in a triplet orbital mixing~\cite{Saraiva2015, Dehollain2014}, to a larger distance limit $\rm{d>8\,nm}$ where the Heitler-London (HL) theory accurately describes the molecular state~\cite{Koiller2001, Wang2016}. The anisotropic envelope plays a role in two ways. First it results in the overall exponential decay of the exchange magnitude with a characteristic length matching the large donor envelope radius $a$. Secondly the anisotropy results in the valley-induced exchange variations to be damped at large distances because of the dominance of the $j_{zz}$ terms (constructive along [110]) over the $j_{xz}$ and $j_{yz}$ oscillating at $\Delta\phi_{[110]}$.

\begin{figure}[htp]
\includegraphics{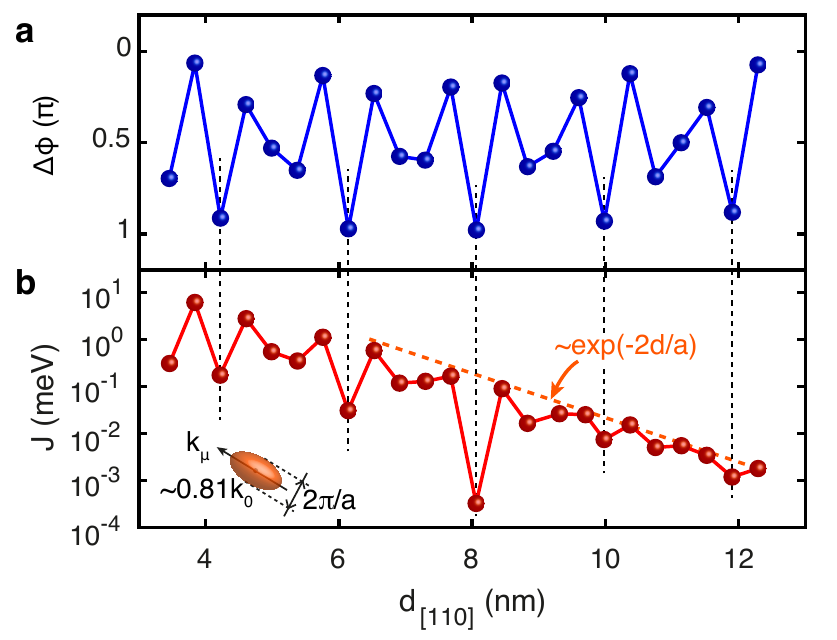}
\caption{ \textbf{Valley interference and exchange interaction.} \textbf{a,} Theoretical phase difference along [110], where $\rm{\Delta\phi_x=\Delta\phi_y=\Delta\phi}$, wrapped between 0 and $\pi$, plotted vs. inter-donor distance along [110]. \textbf{b,} Corresponding exchange value obtained from TB/FCI calculations. $\rm{\Delta\phi}$ and J present a similar oscillating behaviour, with local J minima (resp. maxima), corresponding to valleys being out-of-phase (resp. in-phase), i.e. $\rm{\Delta\phi}$ close to $\pi$ (resp. 0), as indicated by the vertical dashed lines. The red dashed line indicates the exponential decay due the overlap of the wavefunctions envelope parts only, with a characteristic decay length given by the large donor envelope radius $a$.}
\label{figS43}
\end{figure}

\subsection{STM analysis - two-donor position configurations}

We detail here the different donor position configurations when they are meant to be placed along [100] or [110]. The 6 possible positions for each donor results in 36 possibilities for the placement of the two donors, some of them being equivalent. A target along [100] (Fig.~\ref{figS42}a) results in 12 non-equivalent donor-donor position configurations with associated occurrence numbers (adding up to 36) represented in Fig.~\ref{figS42}b, with the convention of only moving $\rm{P_2}$ ($\rm{P_1}$ fixed). Using the same convention, a target along [110] results in 10 non-equivalent configurations represented in Fig.~\ref{figS42}d, also with associated occurrence numbers.\\

\begin{figure}[htp]
\includegraphics{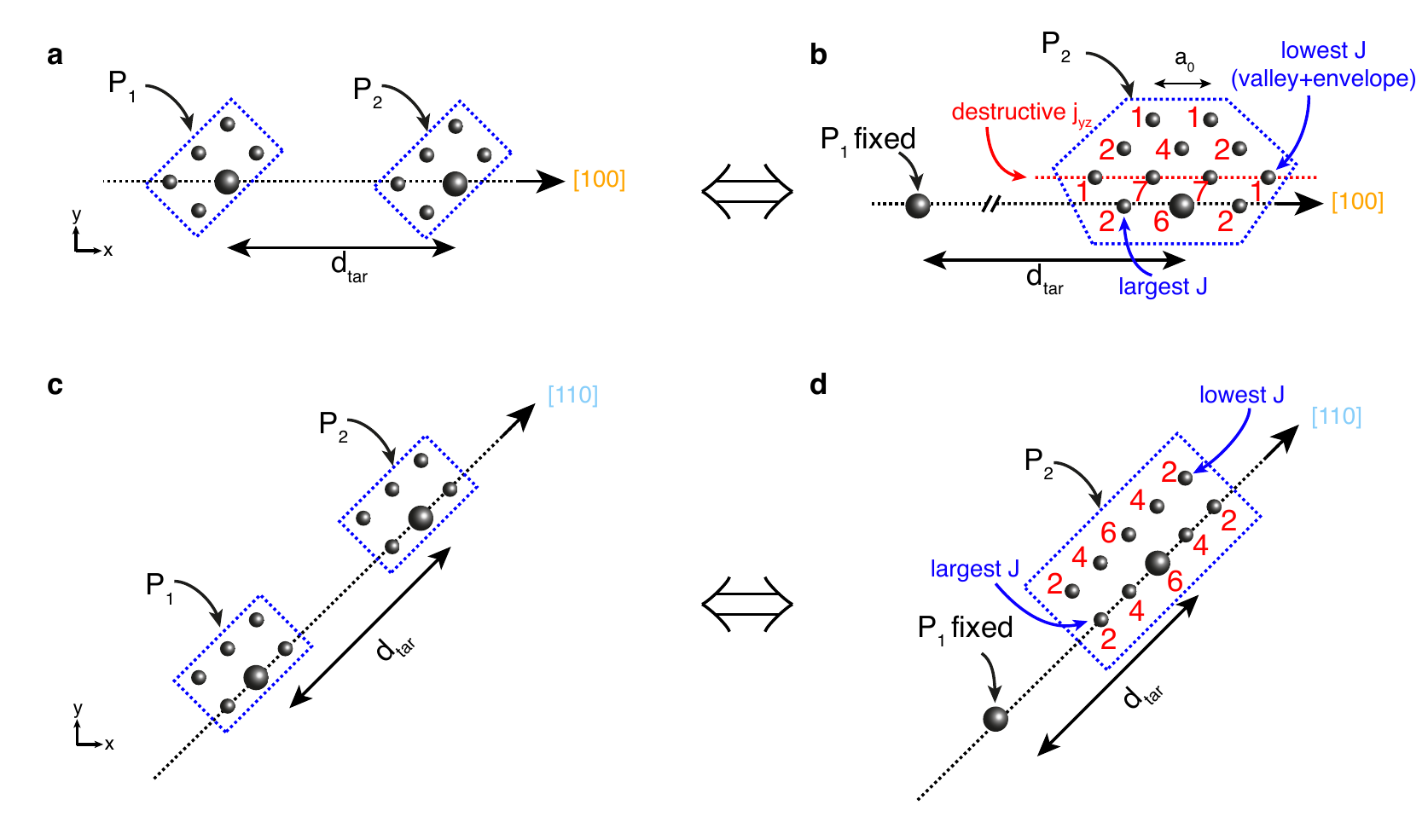}
\caption{ \textbf{Statistical analysis of the two-donor position configurations. a,} Two donors placed along [100], there is 6 possible sites for each of them, i.e. 36 possibilities in total. \textbf{b,} For the purpose of the statistical analysis we can fix the position of $\rm{P_1}$ and work out the 12 possible, non-equivalent, positions for $\rm{P_2}$ as well as their occurrence number (assuming a random distribution of each donor across their 6 possible sites), shown in red, adding up to 36. The destructive positions off by $a_0/2$ along the $y$-axis resulting in low exchange values (see main text) add up to 16 out the 36 possibilities, i.e. 44\% chance to be hit. \textbf{c,} Two donors placed along [110], there is 6 possible sites for each of them, i.e. 36 possibilities in total. \textbf{d,} There are 10 resulting non-equivalent positions for $\rm{P_2}$, assuming $\rm{P_1}$ fixed.}
\label{figS42}
\end{figure}

The P-EM model only uses four parameters: the position of the conduction band minimum $\rm{k_{\mu}}$, the envelope radii $a$ and $b$, which have been measured experimentally, and a normalisation constant $\mathcal{N}$ which can be ignored as we focus on exchange variations and not absolute values. In order to check the relevance of this model we compared it to existing calculations. Gamble $\textit{et al.}$ published a complete set of tunnel coupling $t$ calculations in 3D~\cite{Gamble2015}. We considered all the points in an in-plane slice around a target distance of 12\,nm as shown in Fig.~\ref{figS4}a, squared them to obtain a quantity proportional to exchange (simply assuming $J\sim t^2/E_C$ in the HL limit, where $E_C$ is the charging energy set as constant), normalised them and fitted them using the PEM model. Fig.~\ref{figS4}b shows an exceptional level of agreement, with notably $\rm{k_{\mu}=0.84k_0}$ as used in this reference. Furthermore we empirically calibrated this P-EM model to Heitler-London exchange calculations along [110] based on a tight-binding basis, shown in fig.~\ref{figS4}c. The table shown in Fig.~\ref{figS4}d gives the resulting fitting coefficients obtained for these two fits, with a good agreement between the two theories. The difference in $\rm{k_{\mu}}$ (0.84 against 0.81 in our work) accounts for why the exchange minima are getting out-of-phase with each other at such distances, as observed in ref~\cite{Gamble2015}. The value $a=2.8$\,nm is used in the main text and in this Supplementary Information for any P-EM calculation.

\begin{figure}[htp]
\includegraphics{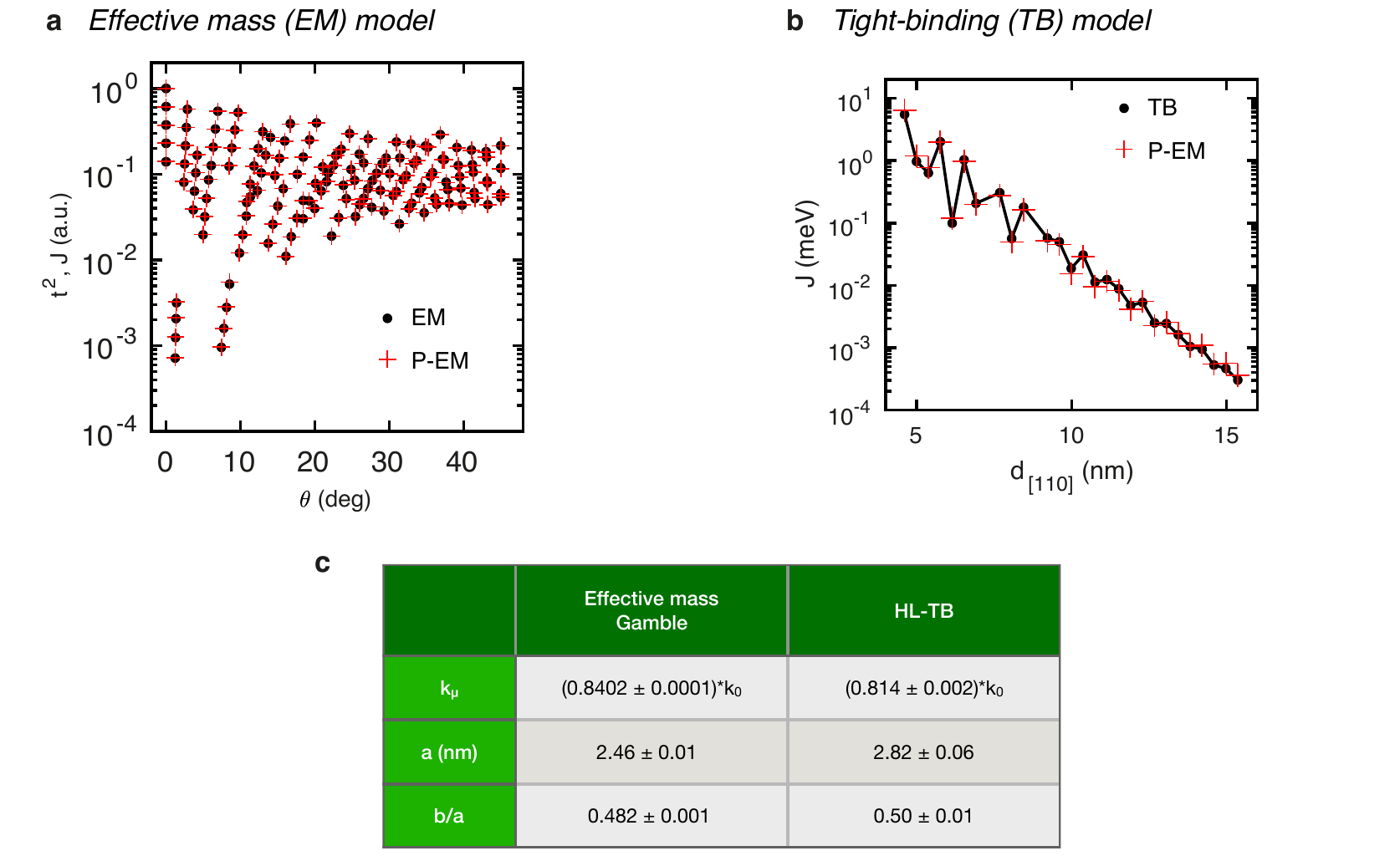}
\caption{ \textbf{Comparison and fit of the P-EM model. a,} Fit to the square of the tunnel coupling values obtained in Gamble \textit{et al.}~\cite{Gamble2015}, for 148 in-plane positions located close to a target distance of 12 nm, between [100] and [110]. \textbf{b,} Fit to the tight-binding calculations developed in this work along [110]. \textbf{c,} Table summarising the fitting parameters obtained in each case. Equations (S18) and (S19) were used to fit both cases, along with a normalisation constant.}
\label{figS4}
\end{figure}

\subsection{Long distance limit}

We can calculate a set of exchange values using the P-EM model for different target distances along both [100] and [110], and extract the minimum and maximum values, respectively $J_{min}$ and $J_{max}$ for each set. We plot in Fig.~\ref{figS52}a the resulting exchange variations, i.e. $\rm{r_{[100]}=J_{min}/J_{max}}$ along [100], and likewise $\rm{r_{[110]}}$ along [110], as a function of $\rm{d_{tar}}$. Along [100] the large variations of more than two orders of magnitude are invariably present due to the destructive locations which are one lattice site off the [100] axis. We can determine the asymptotic limit of $r_{[100]}^\infty$, knowing the donor locations leading to the largest and lowest exchange values (see Fig.~\ref{figS42}), and only considering the $j_{zz}$, $j_{yz}$ and $j_{yy}$ terms (representing 16 out of the 36 terms in total), defining ${d_{tar}=Na_0}$:

\begin{equation}
\begin{gathered}
r_{[100]}=\frac{J_{min}}{J_{max}}\sim\frac{(6+8\cos(\frac{2\pi}{2}0.81)+2\cos(2\pi0.81))\exp(-2a_0/a*\sqrt{(N+1.5)^2+(0.5)^2})}{16\exp(-2(N-1)a_0/a)}\\
\xrightarrow[N \to \infty]{} r_{[100]}^\infty=\frac{3+4\cos(\frac{2\pi}{2}0.81)+\cos(2\pi0.81)}{8}\exp(-5a_0/a) \sim 2.8\times10^{-3}
\end{gathered}
\end{equation}

\noindent assuming $a{=}2.8$\,nm. We have described in the main text the dominance of the $j_{zz}$ terms for donors placed around the [110] direction over the $j_{yz}$ and $j_{xz}$ terms, with a ratio $j_{zz}/j_{yz}$ growing exponentially with distance. In-plane valley-induced variations are hence washed out by the dominance of the $j_{zz}$ terms in the long distance limit. For target distances beyond 12\,nm the variations approach an envelope limit, only set by the distance difference between the shortest and longest inter-donor distances, i.e:

\begin{figure}[htp]
\includegraphics{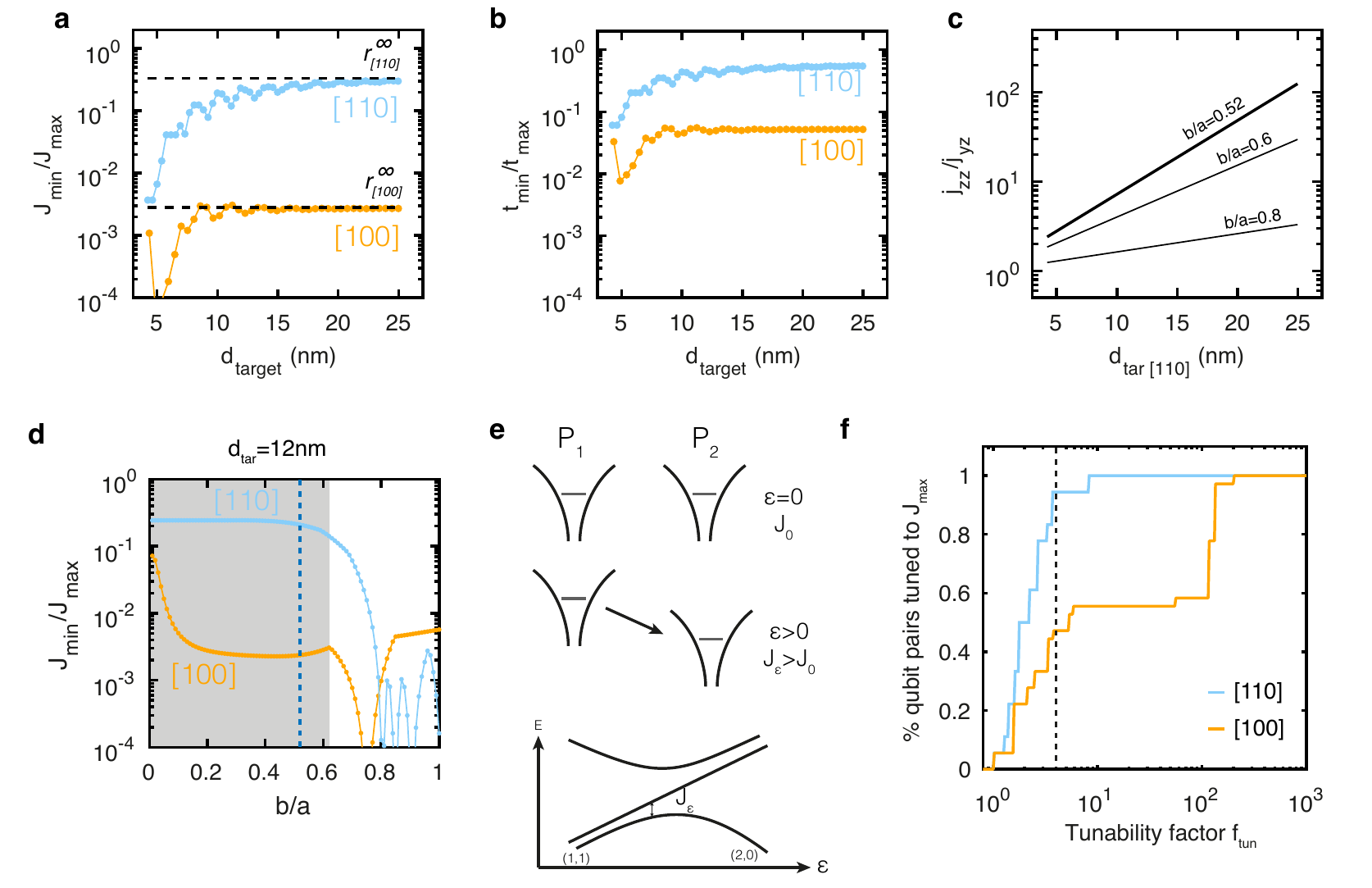}
\caption{ \textbf{Envelope-limited exchange variations and exchange tunability. a,} Plot of exchange variations ${J_{min}/J_{max}}$ and of their asymptotic limit at large $d_{tar}$ of the exchange values obtained from STM dopant placement vs target distances along [100] (orange) and [110] (blue). The variations along [110] reach the envelope limit of a factor 1/0.33, i.e. variations due to the change in inter-donor distance from donor misplacement, while the [100] consistently show more than two orders of magnitude of variations and due to the presence of destructive positions off the [100] axis by $a_0/2$ along $y$. \textbf{b,} Plot of tunnel coupling variations ${t_{min}/t_{max}}$ vs target distances along [100] (orange) and [110] (blue), in the Heisenberg limit valid for inter-donor distances larger than 5\,nm.  The variations along [110] are consistently lower than a factor of 5 for this distance range.\textbf{c} Ratio $j_{zz}/j_{yz}$ plotted vs $d_{tar}$ along [110] for different anisotropy ratios $b/a$. A ratio below 0.6 protects the exchange interaction against in-plane valley interference, i.e. $\Delta\phi_y$ and $\Delta\phi_x$. \textbf{d} Exchange variations plotted vs $b/a$ for $d_{tar}{=}12$\,nm for both [110] and [100] target orientation. The shaded area corresponds to the region where the exchange along [110] is protected against $x$ and $y$-valley interference, i.e. for $b/a{<}0.6$. \textbf{e,} Exchange can be tuned using energy detuning between the two atoms, inducing a (2,0) component in the charge configuration with a larger exchange than the (1,1) configuration. \textbf{f,} Percentage of qubits tuned to the maximum value as a function of a tunability factor.}
\label{figS52}
\end{figure}

\begin{figure}
\includegraphics{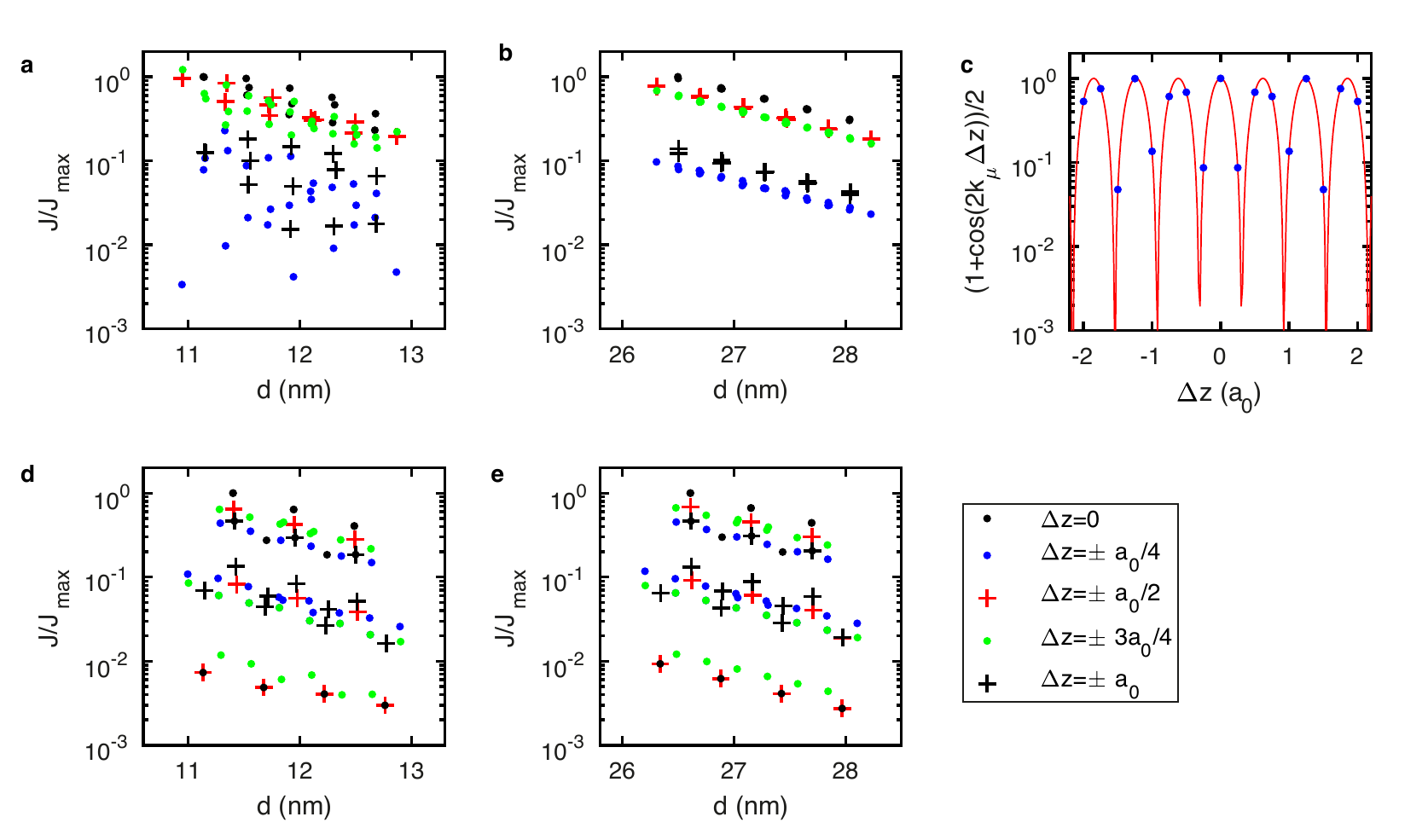}
\caption{ \textbf{Exchange variation analysis including out-of-plane interference. a,} Plot of the exchange values for inter-donor positions found within 1\,nm in-plane circle centred on 12\,nm target distance, and within $\pm1a_0$. An arangement emerges as a function of the $z$-plane difference between the two donors. \textbf{b,} Same for a target distance of 27\,nm. The arrangement is fully set as the exchange only rely on the $z$-valley interference and an envelope function with ${J\sim(1+\cos(2\Delta\phi_z))e^{-2d/a}}$. \textbf{b,} Plot of the $z$-valley interference induced exchange variations ($(1+\cos(2\Delta\phi_z))/2$, red line) for the available inter-atomic plane between the donors (blue spots). \textbf{c,} Same as \textbf{a,} for a 12\,nm target distance along [100]. A complex arrangement can be seen, function of $\Delta\phi_y$ and $\Delta\phi_z$ as the $j_{yy}$, $j_{yz}$ and $j_{zz}$ terms are degenerate. No inter-atomic plane shows a larger exchange variations than that of the in-plane ones discussed in the main tex, which also exist for the $a_0/2$ $z$-plane difference by symmetry. \textbf{d,} Same as \textbf{c,} for a 27\,nm target distance.}
\label{figS5z}
\end{figure}

\begin{equation}
\begin{gathered}
r_{[110]}=\frac{J_{min}}{J_{max}}\sim\frac{\exp(-2a_0/a*\sqrt{(N+1.5)^2+(N+0.5)^2})}{\exp(-2\sqrt{2}(N-1)a_0/a)}\\
\xrightarrow[N \to \infty]{} r_{[110]}^\infty=\exp(-4\sqrt{2}a_0/a) \sim 0.33
\end{gathered}
\end{equation}

The exchange variation analysis we have developed here can be extended to the tunnel coupling, using the relationship $J=4t^2/U$ valid in the Heisenberg regime, i.e. in the weak coupling limit $t{\ll}U$ and $t{\ll}\Delta_{VO}$ (where $\Delta_{VO}$ is the valley-orbit splitting of the order of 10\,meV). This limit corresponds to inter-donor distances larger than 5\,nm~\cite{Klymenko2014,Gamble2015,Saraiva2015}. The tunnel coupling variations can easily be deduced from the exchange coupling variations since $t_{min}/t_{max}=\sqrt{J_{min}/J_{max}}$, plotted in Fig.~\ref{figS52}b vs the target distance along [110] or [100]. We found the tunnel coupling variations are lower than a factor of 5 for any target distance along [110] beyond 5\,nm, as mentioned in the main text.

The asymptotic regime is approached for target distances beyond 12\,nm, for both exchange and tunnel coupling, as shown in Fig.~\ref{figS52}a and b. This a relevant distance range where the exchange coupling has been predicted to be tunable by more than a factor of 5 using electric field detuning~\cite{Wang2016}. This tuning scheme is based on inducing in the ground state a (2,0) charge component (where the two electrons are on the same site), which has a much larger exchange energy than the pure (1,1) state obtained at zero detuning (i.e. zero electric field), as schematised in Fig.~\ref{figS52}b. We plot in Fig.~\ref{figS52}c the percentage of qubits which could be tuned to the same value, i.e. the highest one as this tuning scheme can only increase the exchange, as a function of a tuning factor $f_{tun}$. There are 9 steps for the [110] curve corresponding to the 9 non-equivalent configurations lower than the maximum one, 11 for the [100] curve. A step occurs at a given $f_{tun}$ when a configuration of exchange J reaches the condition $J*f_{tun}=J_{max}$. Each step height reflects the occurrence number of the corresponding configuration, discussed in Fig.~\ref{figS42}. Notably, along [100] the positions leading to destructive $j_{yz}$ and low J values count for 16 out of the 36 possibilities, i.e. almost 50\%, hence large tuning factors (more than 100) are required to tune all the exchange values beyond this threshold. However, along [110] a realistic tuning factor of 4~\cite{Wang2016} is sufficient to tune more than 94\% of the qubit pairs tuned to the same exchange value, reaching the minimum working qubit yield for quantum error correction to be in principle implemented~\cite{Nagayama2017}.

\subsection{$z$-valley interference and exchange}

An important finding of the manuscript is the protection against $x$ and $y$-valley interference along [110]. Here we develop the influence of the $z$-valley interference when the donors are not in the same plane. We have computed the exchange variations according to the P-EM model for all the positions included within a 1\,nm in-plane neighbourhood  and within $\pm1$ monolayer, i.e. $\pm a_0$, for in-plane target distances of 12\,nm and 27\,nm along [100] or [110]. As it is mentioned in the main text, the dependence of the exchange interaction with the $z$-valley interference along [110] is straightforward following the dominance of the $j_{zz}$ terms. In the asymptotic limit, the exchange can be reduced to simply relate to the $z$-valley phase only with ${J\sim(1+\cos(2\Delta\phi_z))e^{-2d/a}}$. This allows to predict exchange variations including possible donor misplacement in the $z$-direction. Since $\Delta\phi_z$ is the only relevant valley phase difference, the exchange variations will be arranged as a function of the depth difference, i.e. atomic planes, between the donors. This was evidenced and discussed in Fig.7 of the main text. This arrangement already emerges for a target distance of 12\,nm as seen in Fig.~\ref{figS5z}a, with an exchange reduced for donor pairs with either a $a_0/4$ or $a_0$ $z$-coordinate difference, and has become asymptotic for a 27\,nm target distance as shown in in Fig.~\ref{figS5z}b. The worst case is for the donors to end up at ${a_0/4}$ depth difference from each other, resulting in variations of $2\Delta J_{[110]}^\infty/(1+\cos(0.81\pi))\sim 41.6$. We plot in Fig.~\ref{figS5z}c the exchange variations due to the $z$-valley interference for a $z$ range within $\pm2a_0$, aligning the possible lattice positions (blue spots) on the resulting variations given by $(1+\cos(2k_{\mu}\Delta z))/2$.~~\\

The exchange variations for target distances along the [100] axis are plotted in Fig.~\ref{figS5z}d and for 12 and 27\,nm target distances respectively. There, an arrangement can also be seen, however it is more complex than the [110] case, as the degeneracy between the $j_{yy}$, $j_{yz}$ and $j_{zz}$ terms results in both $\Delta\phi_z$ and $\Delta\phi_y$ to be involved. This arrangement is already well set up at 12\,nm target distance because the dominance of these terms is even more pronounced than the dominance of $j_{zz}$ along [110] (see Fig.6b of main text). Nevertheless, it appears that the worst case within $\pm1a_0$ variations in $z$ is already contained in the in-plane exchange variations discussed in the main text, which are also present in the $\Delta z{=}a_0/2$ plane by symmetry. This confirm the destructive character of these positions along [100] for the exchange, with the largest variations along [110] including out-of-plane variations being still about one order of magnitude less than the in-plane exchange variations along [100].~~\\

\subsection{In-plane valley interference and exchange along [100] and [110] - Literature comparison.}

In this section, we demonstrate that our results are fully consistent with previous work, and also allow to reconcile their apparent different conclusions. Reference~\cite{Koiller2001} points out to the existence of suppressed exchange values in the close neighbourhood of a target position, reference~\cite{Pica2014} favours the [100] axis for donor placement because of the absence of exchange oscillations along this axis, and reference~\cite{Gamble2015} mentions the absence of a sizeable region where the exchange is stable. References~\cite{Koiller2001} and ~\cite{Pica2014} can be directly compared to our results since they are all based on HL formalism using effective mass wavefunctions, following eq.~\ref{EMenv} with however different parameters summarised in Fig.~\ref{figlit}a. We note that the 3 models have a very similar envelope ratio $b/a$ but reference~\cite{Pica2014} computed a much smaller major Bohr radius (0.9\,nm against 2.5\,nm). The damping of exchange oscillations purely along [100] is related to the ratio between the $j_{xz}$ and the $j_{zz}$ terms. Following the expressions~\ref{envexpress} given above and assuming $x>0$ and $b<a$, this ratio reads:

\begin{equation}
\begin{gathered}
(j_{xz}/j_{zz})(x)_{[100]}\sim \exp(-\frac{x}{a}(\frac{a}{b}-1))\xrightarrow[x \to \infty]{}0
\end{gathered}
\end{equation}

This expression makes clear that the $x$-valley interference impact on the exchange vanishes for a finite anisotropy $b<a$, and is also more pronounced at fixed inter-donor distance along [100] for small $a$~\cite{Pica2014}. Crucially, damping the $x$-valley exchange oscillations also leaves a degeneracy between the $j_{yy}$, $j_{yz}$ and the $j_{zz}$ terms, from which result the suppressed exchange values in the close neighbourhood of [100] because of destructive $y$-valley interference (see main text and "long distance limit" section above). We show in Fig.~\ref{figlit}c-d, the normalised exchange values (with the envelope part taken away in order to focus on the valley interference only) along [100] expected for each model. Both models show over a factor of 100 reduction in exchange for the destructive $y$-positions (see Fig.~\ref{figlit}b), very consistent with both our models shown in Fig.7 of the main text. That is because it is a pure valley interference effect as long as the anisotropy is large enough to create the gap between the $j_{xz}$ and the $j_{zz}$ terms, which we show is largely achieved for these models for very moderate inter-donor distances above 5\,nm.\\

\begin{figure}[htp]
\includegraphics{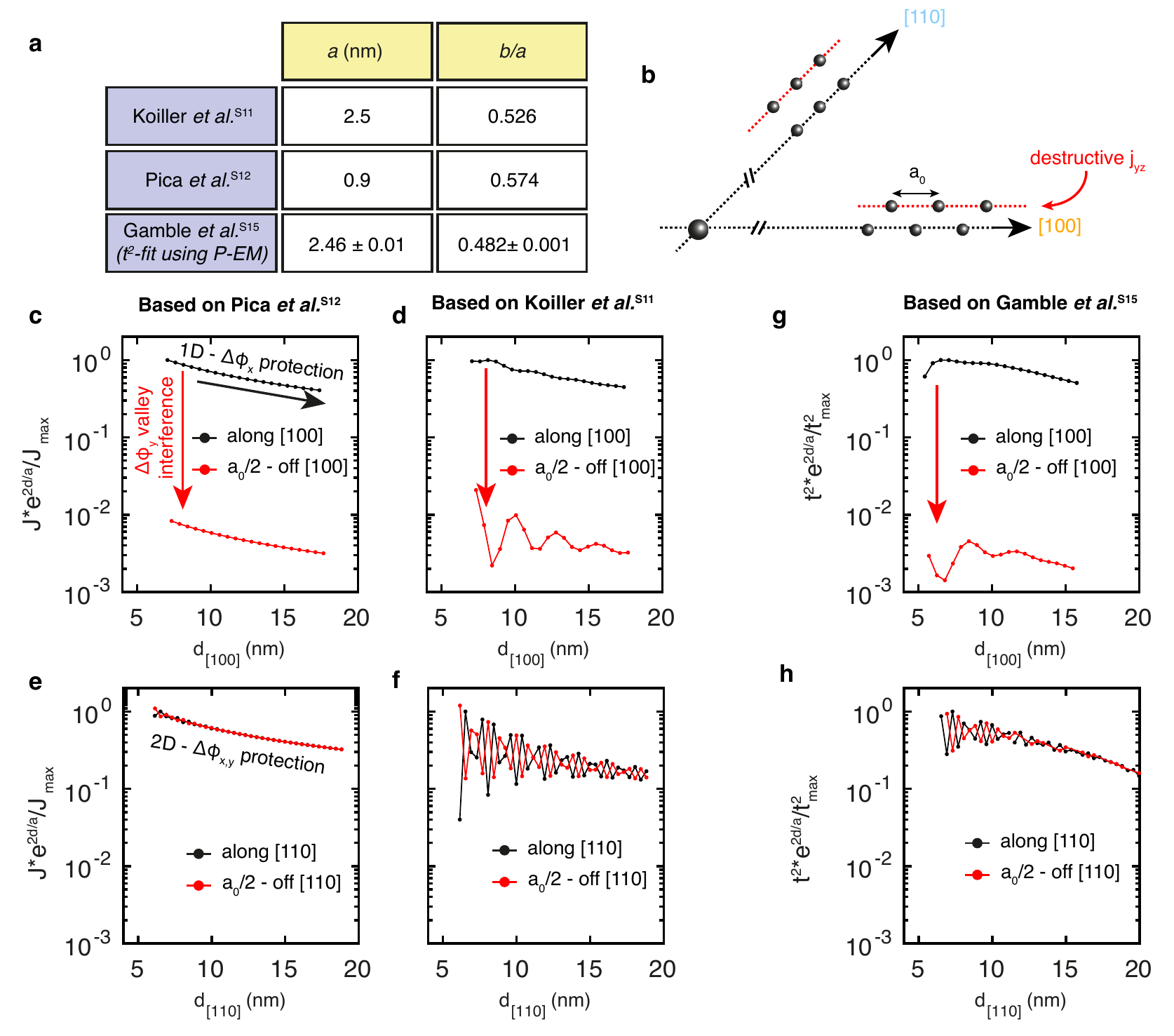}
\caption{ \textbf{In-plane valley interference and exchange along [100] and [110] - Literature comparison. a,} Table summarising the effective mass wavefunctions parameters directly found in reference~\onlinecite{Pica2014} and~\onlinecite{Koiller2001}, and obtained from the fits of the calculations found in reference~\onlinecite{Gamble2015} to our P-EM model. The envelope anisotropy $b/a$ are similar but reference~\cite{Pica2014} found a smaller major Bohr radius $a$ compared to reference~\cite{Koiller2001} and our work. \textbf{b,} Dopant coordinates on and closest off the [100] and [110] axis. \textbf{c,} Normalised HL exchange values  on and off the [100] axis, computed based on reference~\cite{Pica2014} parameters. The exchange envelope decay $\sim\exp(-2d/a)$ was compensated, the remaining decay is due to the Coulomb potential. \textbf{d,} Same for using parameters based on ~\cite{Koiller2001}. For both set of parameters, the exchange is suppressed by more than a factor 100 for the off-[100] positions, which is fully consistent with our results. \textbf{e-f,} Same for the [110], showing that the exchange is protected against in-plane position variations with less than a factor 10 variation, again consistent with our results. \textbf{f-g,} Normalised square of the tunnel coupling calculations directly taken from reference~\cite{Gamble2015}. The envelope decay was also compensated, taking for $a$ the value obtained from the fit to the P-EM model (Fig~\ref{figS4}c). The exchange suppression along [100] as well as the exchange protection along [100] are also very well reproduced.}
\label{figlit}
\end{figure}

Along [110], the ratio between the $j_{xz}$ (or $j_{yz}$ as $y=x$) and the $j_{zz}$ terms along [110] reads:

\begin{equation}
\begin{gathered}
(j_{xz}/j_{zz})(x)_{[100]}\sim \exp(-\frac{\sqrt{2}x}{a}(\sqrt{\frac{a^2+b^2}{2b}}-1))\xrightarrow[x \to \infty]{}0
\end{gathered}
\end{equation}

Exchange oscillations are also damped along [110], however with a smaller rate compared to [100] since $\sqrt{a^2+b^2}{<}2a$. This can visually be seen in Fig.6b of the main text as the gap between the $j_{xz}$ and $j_{zz}$ terms along [100] is larger than that of along [110]. Crucially, along [110] the envelope anisotropy damps both the $x$-valley and the $y$-valley oscillations, hence yielding a 2D exchange protection against valley interference, instead of the 1D protection (as $x$-valley interference are damped only) along [100]. This 2D protection shown in Fig.7d of the main text is well reproduced in Fig.~\ref{figlit}e-f using the parameters found in reference~\cite{Pica2014} and~\cite{Koiller2001}, respectively, with also less than a factor 10 exchange variation. We also show in Fig.~\ref{figlit}f-g that both the strong exchange reduction along [100] and the 2D protection along [110] also features in reference~\cite{Gamble2015} by simply squaring their tunnel coupling calculations, which is also apparent in Fig.\ref{figS4}a.\\

In summary, considering a crystallographic axis for exchange stability with no off-axis displacement makes the [100] direction appealing~\cite{Pica2014} because of the faster decay of the $x$-valley interference impact on exchange, which is enhanced for smaller $a$ values too. But the sensitivity to $y$- and $z$-valley interference along [100] remains, and can be extremely severe when dopants are misplaced from this axis. Instead, [110] results in a higher dimensional protection, which is suitable to the dopant placement accuracy provided by STM lithography. We have shown how considering the interplay between dopant placement, envelope anisotropy and valley interference has allowed to elucidate the origin of the different messages which have been delivered in the literature, and we demonstrate here the universality of our results across different references and models.\\

\newpage

\section{Supplementary Note 4 - Robust two-qubit gates using the exchange interaction}

\begin{figure}[htp]
\includegraphics{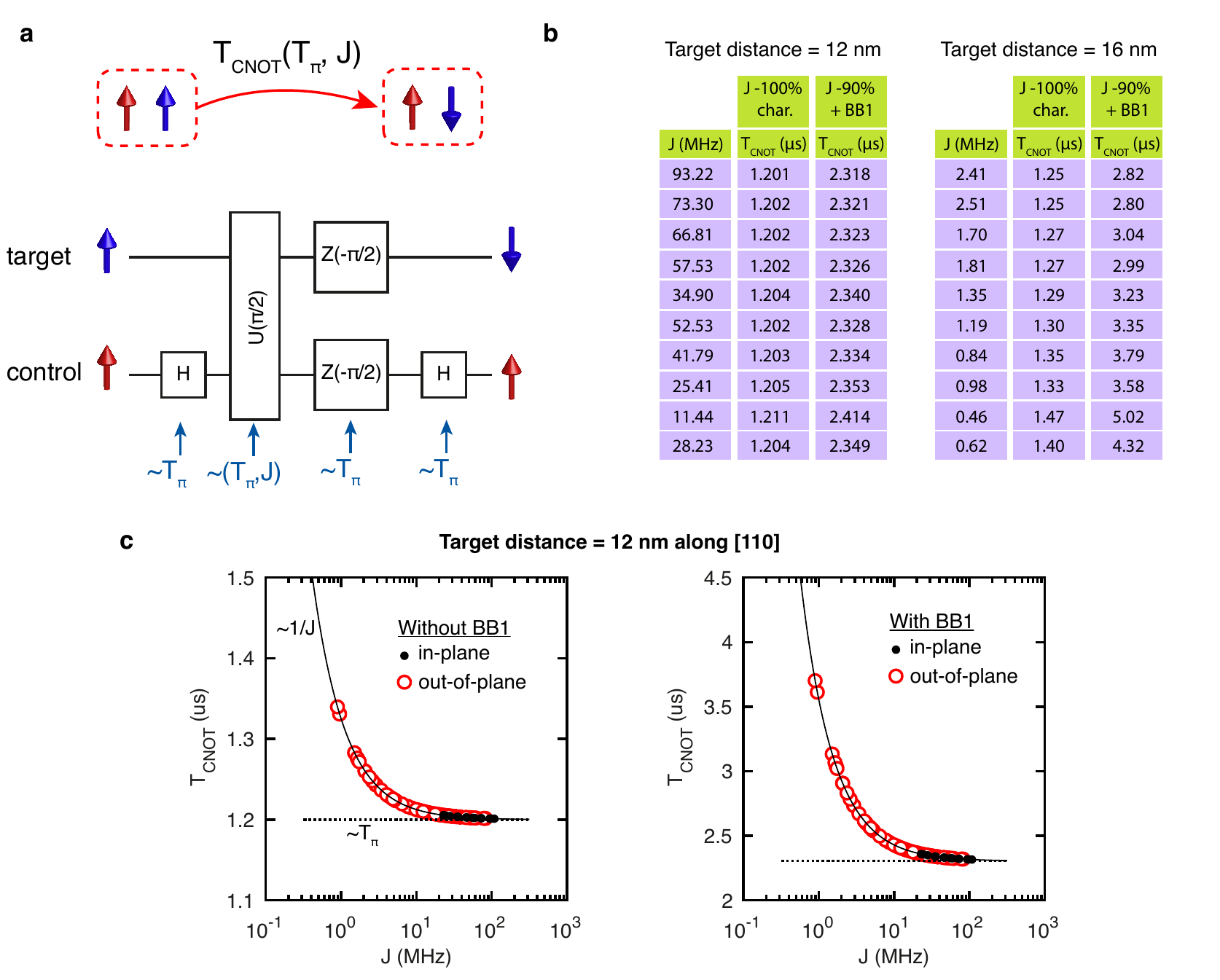}
\caption{ \textbf{Robust CNOT gate based on exchange-coupled donors. a,}. A CNOT gate flips a target spin depending on the sign of a control spin. A composite sequence can achieve beyond 99.9\% fidelity assuming only an exchange characterised to 10\%. It is made of a combination of single qubit rotations, with characteristic time $T_{\pi}$, and of two-qubit rotations, with a characteristic time $h/J$. \textbf{b,} Tables of the calculated CNOT gate times for two target distances along [110], 12\,nm (left) and 16\,nm (right). The 12\,nm histogram shows a very limited spread (1\% without BB1, 4\% with BB1) as the CNOT gate times are limited by $T_{\pi}$ and hence not influenced by small variations in J. The exchange values obtained for a target distance of 16\,nm are lower than the Rabi frequency $1/T_{\pi}$, hence the CNOT gate times are there limited by J and sensitive to exchange variations, accounting for the larger values and spread. It can be compensated by tuning the exchange values as presented in Fig.~\ref{figS52}, bringing more than 94\% of the exchange values (hence CNOT gate times) to the same values. \textbf{c,} Plot of the CNOT gate time values expected for a target distance of 12\,nm, for both without (left) and with (right) the BB1 correction step, for both in-plane and out-of-plane dopant positions according to Fig.~\ref{figS5z}.}
\label{figS5}
\end{figure}

This section describes how a robust CNOT gate based on exchange interaction can be constructed following a sequence of single and two-qubit gates developed in~\cite{Hill2005,Testolin2007}. As shown in Fig.~\ref{figS5}a, this sequence consists of Hadamard gates, i.e. a single qubit $\rm{\pi-rotation}$ around the $\vec{x}+\vec{z}$ axis, $U(\pi/2)$ a two-qubit rotation and $Z(-\pi/2)$ single qubit $\rm{\pi/2}$-rotation around the Z-axis. The calculated CNOT gate times are given in the tables shown in Fig.~\ref{figS5}b, for two target distances 12 and 16\,nm along [110]. As mentioned in the main text, for 12\,nm the exchange values range between 93 and 11\,MHz, i.e. all larger than the Rabi frequency, fixed at 3\,MHz~\cite{Pla2012}. This target distance represents a good compromise between benefiting from the [110] exchange stability scheme and from large enough exchange values for the CNOT gate times to be mainly limited by $T_{\pi}$. The spread in CNOT gate time expected for the in-plane configurations is limited to 1\% around a mean value of 1.2\,$\mu$s. Considering out-of-plane configurations within 1 monolayer, the maximal CNOT gate time is 1.34\,$\mu$s (see Fig.~\ref{figS5}c). The BB1 sequence compensates for rotation errors if the exchange values are characterised to only 90\% accuracy experimentally, to keep fidelities above 99.9\%. In this case, in-plane CNOT gate times increased to an average value of 2.3\,$\mu$s with a spread limited to 4\%. The expected BB1 CNOT gate time for the out-of-plane configurations can reach up to 3.7\,$\mu$s for the lowest exchange values, however we note that only 6 out of 82 computed configurations led to an operation time above 3\,$\mu$s. This means that for this target distance, a factor of 100 variations in exchange results in less than a factor of 2 in the CNOT gate time. We performed similar calculations and analysis for two donors placed at 16\,nm along [110]. For this target distance, the exchange values range between 2.4 and 0.5\,MHz, which is comparable or lower than the Rabi frequency. As a consequence, the spread in exchange has more impact on the CNOT gate times, ranging now between 1.25 and 1.40\,$\mu$s (up to 3.5\,$\mu$s for out-of-plane configurations), and between 2.8 and 5.0\,$\mu$s (up to 25.4\,$\mu$s for out-of-plane configurations) if a BB1 step is involved. The in-plane spreads can be compensated by electrically tuning the exchange values as presented in the previous section, bringing more than 94\% of the qubit pairs to the exact same exchange value with a realistic tuning factor of 4~\cite{Wang2016}.~~\\

\end{widetext}


\end{document}